\documentclass[
 a4paper, twocolumn, aps, prb, longbibliography, superscriptaddress, floatfix]{revtex4-2}

\usepackage{xcolor}
\usepackage{graphicx}	
\graphicspath{ {images/} }					
\usepackage{amsmath} 						
\usepackage{bbold}							
\usepackage{physics} 						
\usepackage{amssymb}						
\usepackage{dsfont}                         
\usepackage{breqn}							
\renewcommand{\vec}{\vectorbold}          

\definecolor{GraphRed}{HTML}{A31621}
\definecolor{GraphBlue}{HTML}{15616D}
\definecolor{GraphGreen}{HTML}{88498F}

\newcommand*\dif{\mathop{}\!\mathrm{d}}		
\newcommand*\bigO{\mathop{}\!\mathcal{O}}   

\newcommand{\figref}[1]{\figurename~\ref{#1}}

\newcommand{\Lhomo}{L_{\mathrm{hom}}}
\newcommand{\Ldyn}{L_{\mathrm{dyn}}}

\begin{document}   
\date{\today}
\title{Elastic properties and thermodynamic anomalies of supersolids} 
\author{Milan Rakic}
\affiliation{Department of Physics\char`,{} Blackett Laboratory\char`,{}
 Imperial College London\char`,{} SW7 2AZ\char`,{} UK}
\author{Andrew F.~Ho}
\affiliation{Department of Physics\char`,{} Royal Holloway University of
London\char`,{} Egham\char`,{} Surrey TW20 0EX\char`,{} UK}
\author{Derek K.~K.~Lee} 
\affiliation{Department of Physics\char`,{} Blackett Laboratory\char`,{}
 Imperial College London\char`,{} SW7 2AZ\char`,{} UK}

\begin{abstract}
    We study a supersolid in the context of a Gross-Pitaevskii theory with a
    non-local effective potential. We employ a homogenisation technique which
    allows us to calculate the elastic moduli, supersolid fraction and other
    state variables of the system. Our methodology is verified against numerical
    simulations of elastic deformations. We can also verify that the
    long-wavelength Goldstone modes that emerge from this technique agree with
    Bogoliubov theory. We find a thermodynamic anomaly that the supersolid does
    not obey the thermodynamic relation $\partial P / \partial V |_N \, = - n \,
    \left( \partial P / \partial N |_V \right)$, which we claim is a feature
    unique to supersolids. 
\end{abstract} 

\maketitle

\section{Introduction} 
A supersolid is a phase of matter that displays both crystalline order and
superfluidity in the form of non-classical rotational inertia (NCRI). A key
requirement is that both the continuous translational symmetry and the U(1)
global gauge symmetry of the system is spontaneously broken in the ground state. 

There have been several attempts to understand this quantum phase of matter
experimentally and to understand its properties theoretically. While it has not
been observed in bulk \textsuperscript{4}He \cite{kim2014upper,day2007low},
there are hints that such a phase may exist in the second monolayer of
\textsuperscript{4}He on graphite \cite{nyeki2017intertwined} where NCRI has
been measured in a density regime near layer completion with an anomalous
temperature dependence of the specific heat capacity. More recent experiments in
ultracold dysprosium and rubidium atoms
\cite{chomaz2019long,guo2021optical,norcia2021two} have observed the spontaneous
breaking of continuous translational symmetry together with long-range phase
correlation. 

The first considerations of a supersolid phase were made by Andreev and Lifshitz
\cite{andreev1969quantum} and Chester \cite{chester1970speculations} who
considered the possibility of a supersolid phase in \textsuperscript{4}He. They
argued that the superfluid fraction of a supersolid would be reduced from 100\%
due to the coupling of the phonons of the crystalline
structure to the $U(1)$ phase of the condensate wavefunction. This was further
developed by Leggett \cite{leggett1970can}.

Later attempts at theoretical work have taken a phenomenological symmetry-based
approach starting with the work of Nozi\`eres \cite{nozieres2009superfluidity}
and Dorsey et al.~\cite{dorsey2006squeezing}. Son \cite{son2005effective} has
described how Galilean invariance puts constraints on the form of the
Lagrangian that describes the low-energy dynamics of supersolids.

There have also been approaches starting from microscopic Hamiltonians based on
Gross-Pitaevskii theory \cite{pomeau1994dynamics} and Bogoliubov theory
\cite{kunimi2012bogoliubov}. It is natural to ask whether these approaches give
rise to the same predictions for the low-energy properties of the supersolid
system, such as elasticity and superfluidity.

Moreover, there has been intensive study on the superfluid fraction and
excitation spectrum of the supersolid phase. However, there has been
comparatively little study on the elastic properties. In this paper we will
build on the homogenisation technique of Rica and co-workers
\cite{josserand2007patterns,josserand2007prl,during2011theory} to obtain an
effective low-energy theory that agrees well with Bogoliubov theory and
numerical calculations. Importantly, the original formulation of
the homogenisation technique obtained a value for the bulk modulus that was not
in agreement with the expected bulk modulus in the superfluid phase.
Furthermore, a triangular supersolid would be expected to possess certain
symmetries in the Cauchy elastic tensor, and these were not satisfied in the
original formulation. Finally, the velocities of the long-wavelength excitations
in the previous work also did not agree with the results based on Bogoliubov
theory \cite{kunimi2012bogoliubov}, in the sense that the velocities obtained on
a particular supersolid phase using the Bogoliubov technique were not consistent
with those obtained via the original formulation of homogenisation.

With some important corrections that will be derived in this paper, we will show
that the homogenisation approach can be reconciled with other techniques,
producing the predictions for all the elastic and superfluid properties of the
supersolid phase. More specifically, we provide a method with the
Gross-Pitaevskii approximation that calculates the elastic constants and
superfluid density (phase stiffness) which is valid for any particular
ground-state solution of a Gross-Pitaevskii Lagrangian. It should be noted that,
in the limit of zero temperature, where a homogeneous superfluid has a
superfluid fraction of 100\%, a supersolid has a reduced superfluid fraction
\cite{leggett1970can,leggett1998superfluid}. It is important to clarify that
this is not an enhancement of the ``normal fraction'' which is found in
superfluids at non-zero temperatures and carries entropy. Rather, it is
indicative of a separate phase that spontaneously breaks translational symmetry.
In fact, it is a consequence of the fact that the ``phonons'' of the crystalline
structure couples to the $U(1)$ phase of the condensate wavefunction.
\cite{sepulveda2008nonclassical,
roccuzzo2020rotating,aftalion2007nonclassical,tanzi2021evidence,
blakie2023superfluid}. 

The outline of the paper is as follows. In section 2, we review the
Gross-Pitaevskii theory for Bose condensates and how a finite-range interaction
can give rise to a supersolid phase. In section 3, we lay the framework for the
homogenisation theory, specifying the deformation procedure and rigorously
defining all necessary steps to calculate a long-wavelength theory of the
low-lying excitations. Section 4 uses results from section 3 to derive elastic
constants analytically and finds that they agree with elastic constants that we
obtain numerically. In section 5, we will consider the additional $U(1)$ gauge
and derive the supersolid fraction, as well as a coarse-grained
Lagrangian which is the analytic expansion around the ground-state of the
elastic strain and $U(1)$ fields. Section 6 uses the effective Lagrangian and
considers the additional energetic contributions of flow across the system, and
of work done by the strain on the surroundings. In doing so, we derive a
long-wavelength effective Lagrangian which describes the Goldstone modes of the
system. Section 7 solves for the excitation velocities and discusses some of the
key features of the theory. In section 8, we verify the excitation velocities
through use of Bogoliubov fluctuations in a Bloch theorem context, and find
excellent agreement between the two seemingly disparate theories. In Section 9, 
we will discuss a thermodynamic anomaly in the compressibility that we believe
is unique to a supersolid.

We subsequently make the claim that since we have satisfied the symmetry
requirements for an effective theory imposed by Son \cite{son2005effective},
have verified our elastic constants numerically, and have verified the
velocities of the low-lying excitations through two independent and seemingly
unconnected techniques; then we must have the correct effective Lagrangian for a
supersolid. Furthermore, the technique outlined in this paper can now be used
and applied directly to other more complicated and realistic systems. 

\section{Meanfield Supersolids} 

In this paper, we study a Bose-Einstein condensate (BEC) of particles of mass
$m$ with a finite-range interaction $U(\mathbf{r})$ in a Gross-Pitaevskii
theory. The condensate wavefunction $\psi(\mathbf{r},t)$ is a complex-valued
function that can be written in number-phase representation (or Madelung form)
as $\psi(\mathbf{r},t) = \sqrt{\rho(\mathbf{r}, t)} e^{i \phi(\mathbf{r}, t)}$.
The Lagrangian of the system is 
\begin{equation}\label{EQ:GP Lagrangian Madelung}
    \begin{split}
        \mathcal{L} = - \int_{\Omega} & \left[ \hbar \rho \frac{\partial\phi}{\partial t} + \frac{\hbar^2}{2m} \left(
        \rho (\nabla \phi)^2 + \frac{1}{4 \rho} (\nabla \rho)^2 \right) \right. \\
        & \left. + \frac{1}{2} \, \rho (\mathbf{r})  \int_{\Omega} U(\abs{\vec{r} - \vec{r}'}) \rho(\vec{r}') \dif \vec{r}' \right] \dif \vec{r} 
    \end{split}
\end{equation}
where $\Omega$ is the spatial domain of the system.  
In the absence of any current or twisted boundary conditions, the phase $\phi$
is spatially uniform in the ground state and can subsequently be set to zero.
The ground state density
can be found by variational calculus and is given by 
\begin{equation}\label{EQ:rho base EoM}
    \frac{\hbar^2}{4m} \left( \frac{(\nabla \rho)^2}{2 \rho^2} 
    -  \frac{\nabla^2 \rho}{\rho} \right) 
    + \int_\Omega U(\vec{r} - \vec{r}') \rho(\vec{r}') \dif \vec{r}' = \mu \, .
\end{equation}
where $\mu$ is the chemical potential to enforce the constraint $\int_\Omega \rho(\mathbf{r}) \dif
\mathbf{r} = N$, the total number of particles in the system
Equation \ref{EQ:rho base EoM} is a nonlinear equation which we solve
numerically via a Crank-Nicholson scheme. By evolving the Lagrangian in
imaginary time numerically, we obtain the (real) ground state wavefunction as
the steady-state solution, using a Runge-Kutta algorithm for the temporal
evolution. The numerics are performed on a 2D triangular grid in real space with
periodic boundary conditions in order to respect the hexagonal symmetry of the
spatially modulated condensate that we expect.

A necessary feature (as shown by Heinonen et. al \cite{heinonen2019quantum}) for
the spontaneous development of a density wave in the condensate is the presence
of an interaction potential which at least has non-zero Fourier components in
$\mathbf{k}^4$ and upwards \cite{heinonen2019quantum}. This rules out the use of
the typical contact potential for supersolid formation. 

To illustrate this, consider a toy model of a soft-core repulsive interaction
with a finite-range of $a$ in two dimensions: 
\begin{equation}
    U(\vec{r}) = U_0 \Theta(\abs{\mathbf{r}} - a)
\end{equation}
where $\Theta$ is the Heaviside step function. For this simple interaction, the
system is controlled by a single dimensionless parameter 
\begin{equation}
    \Lambda = \frac{\pi U_0 m a^2}{\hbar^2} n a^{2} 
\end{equation}
for a system of average number density $n$.  

\begin{figure}[htb]
    \includegraphics[width=\columnwidth]{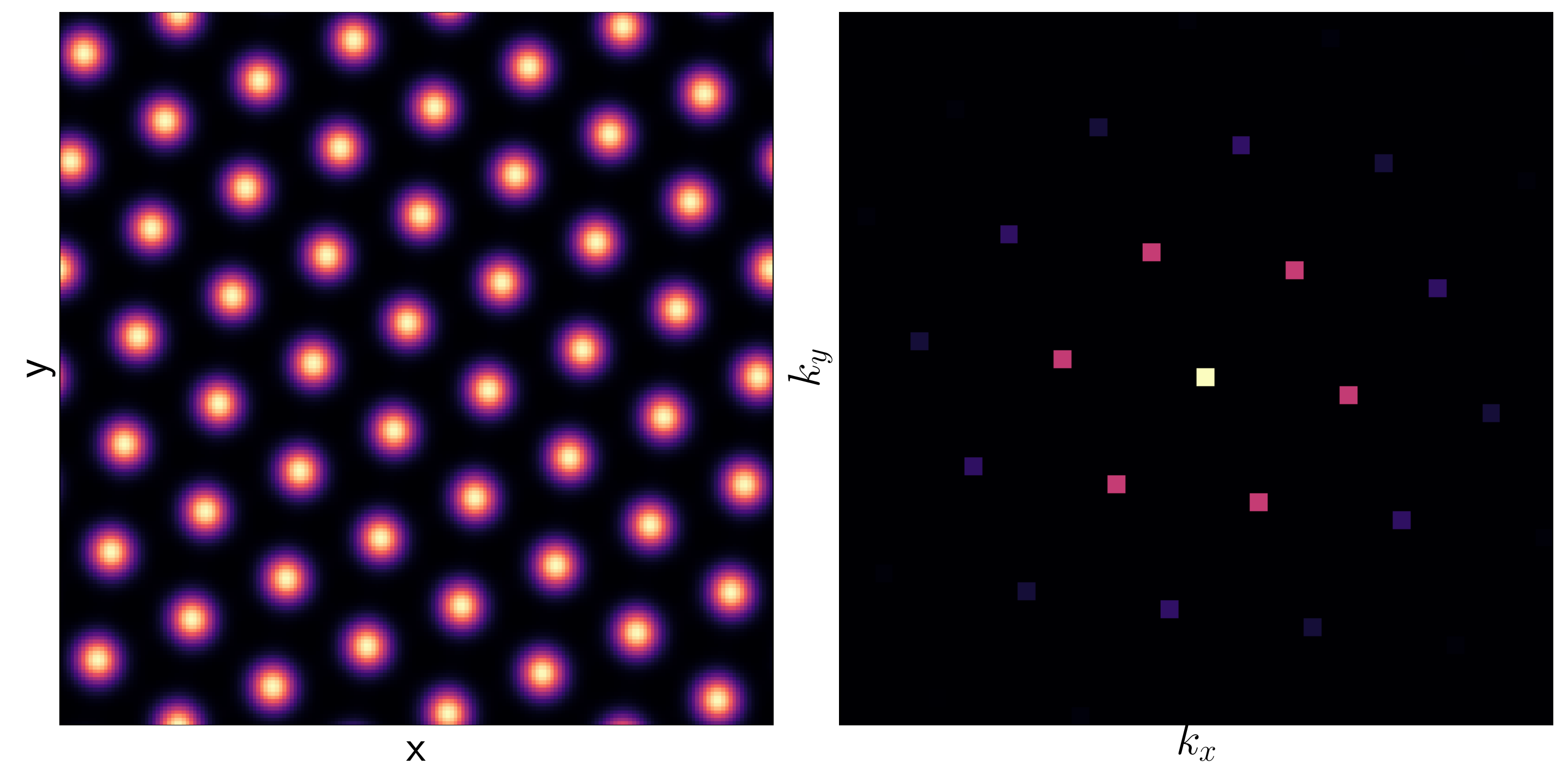}
    \caption{A spatially modulated condensate. Left: real space density
    distribution. Right: Fourier transform of density distribution implying
    non-zero momenta in the condensate.}
    \label{FIG:supersolid phase real and fourier} 
\end{figure}

The phase diagrams for the soft-core system are described in detail in various
works
\cite{wessel2005supersolid,boninsegni2005supersolid,boninsegni2012supersolid}.
We find that our results are in agreement, with small $\Lambda$ corresponding to
a superfluid phase and that for $\Lambda > 37$ the system spatially orders into
the supersolid phase. This is in agreement with a theoretical subcritical limit
calculated by During et al \cite{during2011theory}. This supersolid phase is
characterised by a triangular lattice in 2D, illustrated in
\figref{FIG:supersolid phase real and fourier}. 

The bulk of this paper is concerned with an effective long-wavelength theory for
this supersolid phase where the Bose condensate is spatially modulated. We will
use these numerics to verify our formalism by calculating the elastic constants
which are dependent on the ground state. 

\section{Homogenisation for Elasticity}\label{sec:homogenisation}

The spontaneously broken translational symmetry in a supersolid means that the
system must possess Goldstone modes which are the elastic modes. In this
section, we will build on the work of
\cite{josserand2007patterns,during2011theory} to derive the elastic response of
the system. The goal here is to consider an elastic deformation on the ground
state, e.g. stretching/shearing as described by some strain tensor, and then to
expand the Lagrangian to second order in said strain tensor and recover the
elastic moduli through the generalised Hooke's law. 

We follow the metholodology of Josserand et al \cite{josserand2007patterns}, but
our results have some key differences. Namely, we believe some key terms
were missed which skews the elastic constants to ones that do not agree with
simple calculations in the superfluid limit and lead to physically inconsistent
elastic moduli. More specifically, the Cauchy elastic tensor does not obey
internal symmetries consistent with a triangular lattice, and possesses negative
elastic moduli which suggest an unstable state. Moreover, we conduct a different
treatment of the system with regard to the canonical ensemble and the degrees of
expansion in the strain tensor.

The analysis will take the following structure: we will first carefully define
the notion of an elastic deformation on the system and provide the basis for the
calculation. We then expand the Lagrangian to second-order in the elastic strain
tensor. We demand that the deformed system also be a ground state, and therefore
solve the Euler-Lagrange equations for the new system, which will allow us to
determine elastic constants. 

We begin by describing a physical deformation by considering
the displacement, $u(r)$, of points in the material. The
displacement can be written as 
\begin{equation}\label{EQ:deformation}
    \mathbf{r}' = \mathbf{r} - \mathbf{u}(\mathbf{r}) \, , 
\end{equation}
moving points $\mathbf{r}$ (over some domain $\Omega$) in the undeformed ground
state to points $\mathbf{r}'$ (with some domain $\Omega'$). This basis describes
the deformed material in the lab frame of an external observer. To clarify, this
is considered an \textit{active} transformation which physically acts on the
material and changes the domain of the system in the lab frame. The basis
$\mathbf{r}'$ is a Cartesian coordinate system in the lab frame of the external
observer, but is a non-Cartesian coordinate system in the frame of the material.

We then define an inverse \textit{passive} transformation which maps the
coordinate system $\mathbf{r}'$ to the coordinate system $\mathbf{r}''$, which
is a non-Cartesian basis in the material frame that has the same domain as the
inital $\mathbf{r}$ (domain $\Omega$). This passive
transformation is defined by the displacement $\vec{u}' ( \vec{r}' )$, where we
note that this does not change the energy of the system in any way. It is simply
a redefinition of coordinates in order for us to express the actual change in
the energy that occured during the \textbf{active} transformation defined in
\eqref{EQ:deformation}.

This passive transformation is given by the form 
\begin{equation}\label{EQ:inverse deformation}
    \mathbf{r}'' = \mathbf{r}' + \mathbf{u}'(\mathbf{r}') \, , 
\end{equation}
where the basis $\mathbf{r}''$ coincides identically with the basis $\mathbf{r}$
in the material frame, i.e. $\mathbf{r}'' = \mathbf{r}$, but now the space has
additional components in its gradients, as well as a Jacobian associated with
its volume element. Note that the passive transformation $\vec{u}'(\vec{r}')$ is
in the deformed lab frame given by the coordinates $\vec{r}'$, and importantly
is a different transformation to $\vec{u}(\vec{r})$.

\begin{figure}[h]
    \includegraphics[width = \linewidth]{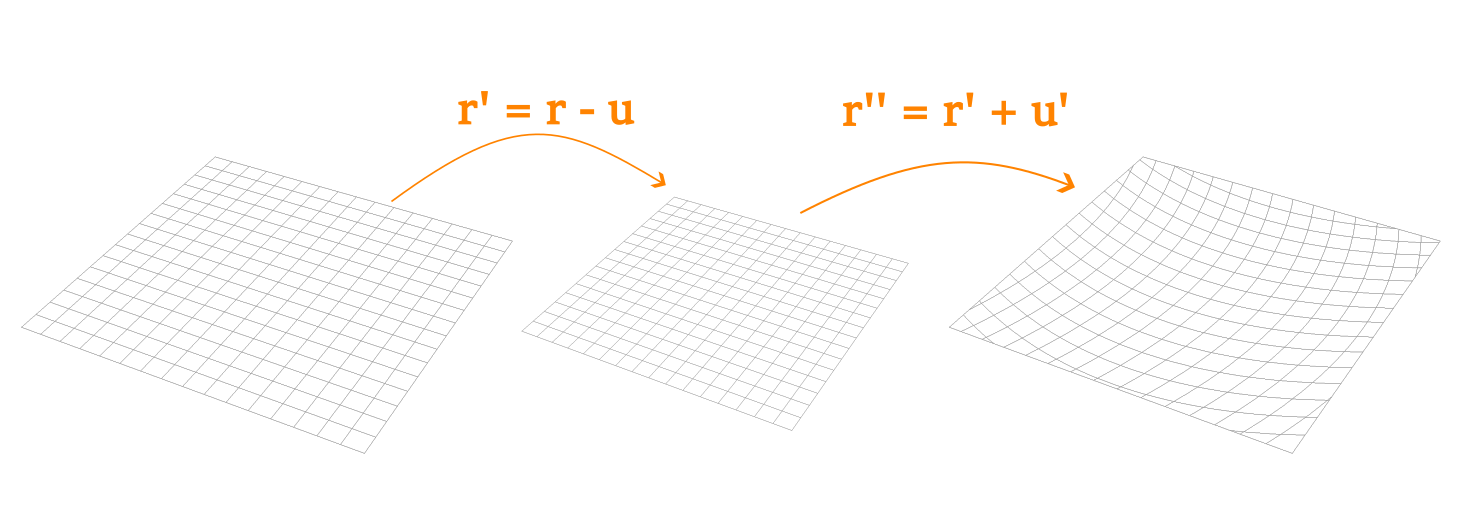}
    \caption{A schematic of the deformation procedure. The leftmost image
    depicts an undeformed system with a basis $\mathbf{r}$, which then undergoes
    an \textit{active} transformation given by some strain to become the middle
    system. The rightmost image depicts the system after the \textit{passive}
    transformation, such that the final coordinate system $\mathbf{r}''$
    coincides exactly with $\mathbf{r}$, but now there are additional curvatures
    in the space. }\label{FIG:deformation schematic}
\end{figure}

These extra curvatures are necessary to describe the full
effect of the displacement procedure on the system. We now relabel the basis
$\mathbf{r}''$ as $\mathbf{r}$, for convenience, where again we must stress that
even though the coordinate systems $\mathbf{r}$ and $\mathbf{r}''$ coincide, the
space is no longer Cartesian and has an associated Jacobian and curvature. This
procedure is illustrated schematically in \figref{FIG:deformation schematic}. 

As of now, the above transformation is completely general, but in order to
proceed we need to specify a particular class of transformation which allows us
to use linear response theory and elasticity theory. We make the distinction
that $\vec{u}({\vec{r}})$ is such that all gradients in space are small
constants and that gradients in time are considered small velocities.

More specifically, we choose the tensor $\partial_i u_k$,
which is a dimensionless number, s.t. $\partial_i u_k \sim \delta$, where
$\delta \ll 1$. The time derivative of $\vec{u}$, i.e. $\partial_t u_k$, is
treated as small in the same sense as the phase perturbation, which we will
later cover in Section \ref{sec:phasegradient}. This is referred to as a uniform
deformation, in that the strain tensor is constant and small everywhere, which
allows us to treat it as a perturbative parameter which we will expand up to
harmonic terms.

We now refer to $u_{ik} \equiv \partial_i u_k$ as the \textit{strain tensor},
and endeavour to expand $\mathcal{L}$ in powers of $u_{ik}$. In order to employ
linear elastic theory we have to keep all terms up to $\bigO(\delta^2)$ in the
Lagrangian. We also need to relate the gradient of the transformation $\vec{u}'$
in the primed deformed basis in the lab frame to the gradient of the
transformation $\vec{u}$ in the material basis. It is relatively easy to show
that 
\begin{equation}
    \partial_i' u_k' = \partial_i u_k + \partial_i u_l \partial_l u_k 
    \equiv u_{ik} + u_{il}u_{lk} 
\end{equation}
up to $\bigO(\delta^2)$ in the strain tensor (n.b. we are using the Einstein
summation convention). This allows us now to express derivatives in the primed
(deformed lab frame) basis in terms of derivatives in the unprimed (deformed
material frame) basis to find 
\begin{equation}
    \partial_i' = \partial_i + \partial_k \left( u_{ik} + u_{il} u_{lk} \right) \quad.
\end{equation}

In order to simplify the expressions that we will derive in this paper, we
introduce notation for a set of key tensors that appear regularly. 
\begin{equation}\label{brevity tensor definitions}
    \begin{split} 
        &\epsilon_{ik} = -\frac{1}{2} \left( u_{ik} + u_{ki} \right) \quad ; \quad \Delta_{ik} = \frac{1}{2} u_{il} u_{kl} \quad ; \\  
        &\omega_{ik} = \frac{1}{2} u_{li} u_{lk} \quad ; \quad \chi_{ik} = \frac{1}{2} \left( u_{li} u_{kl} + u_{lk} u_{il} \right)
    \end{split}
\end{equation} 

Since the deformation actively changes the domain of the system, there is an associated change of volume. We refer to the deformations described in
\eqref{EQ:deformation} and \eqref{EQ:inverse deformation} to define the
differential volume elements as 
\begin{equation}
    \int_{\Omega'} \dif \vec{r}' = \int_{\Omega} \mathcal{J}_{\vec{r}' \rightarrow \vec{r}} \dif \vec{r} 
\end{equation}
with the Jacobian 
\begin{equation}\label{jacobian volume expansion}
    \mathcal{J}_{\vec{r}' \rightarrow \vec{r}} = \det(\frac{\partial r_i
    '}{\partial r_k}) = 1 + \epsilon_{ll} + M_{iklm} \, u_{ik} u_{lm} \quad. 
\end{equation}
where $M_{iklm}\equiv
(\delta_{ik}\delta_{lm} - \delta_{im}\delta_{kl})/2$. 
A negative $\epsilon_{ll}$ corresponds to a reduction in volume. In two dimensions, $M_{iklm}u_{ik}u_{lm} = u_{xx}u_{yy} - u_{xy}u_{yx}$. The quantities $\epsilon_{ll}$ and $M_{iklm}u_{ik}u_{lm}$ are sometimes called the first and second strain invariants as they are independent of the basis in which the strain tensor is written. We stress now that the
coordinate system $\mathbf{r}$ describes the deformed system in the frame of
the material: it has the same domain as the undeformed material but is now a
non-Cartesian space.

To understand how this deformation affects the Lagrangian, we need to understand
how the density changes as a function of the deformation. After the deformation
occurs, the particles in the system will reorganise themselves in such a way as
to minimise the total energy of the new configuration, leading to a new density
described by $\rho(\mathbf{r}')$. The reorganisation of particles due to a
deformation can be expressed as the pre-deformation density plus a component,
$\tilde{\rho}$, which completely accounts for all changes. The component
$\tilde{\rho}$ covers both local density changes at length scales within a unit
cell of the supersolid and also includes the change in average density due to
total number conservation and a change in volume.

We consider that this new density is written in the $\mathbf{r}'$ basis. We
stress that this is not the density in the material frame, but rather the
density in the lab frame of the external observer. We can then use the passive
transformation to write the density in the $\mathbf{r}''$ basis, which is the
material frame that now has some curvature. Without loss of generality we can
express this new density at some point $\mathbf{r}'$ as 
\begin{equation}\label{EQ:density analytic expression}
    \begin{split}
        \rho(\mathbf{r}') &= \rho_0(\mathbf{r}'') + \tilde{\rho}(\mathbf{r}'')\,, 
        \int_{\Omega'} \!\!\!\rho(\mathbf{r}') \dif \mathbf{r}' 
        = \int_\Omega\!\!\!\rho_0(\mathbf{r}'') \dif \mathbf{r}''
    \end{split}
\end{equation}
where $\rho_0$ is the pre-deformation ground-state density profile. As of now
this is simply a mapping to a scalar function that has been subject to an active
transformation. 

We stress that from this point we drop the $\mathbf{r}''$ notation and refer to
the coordinate as $\mathbf{r}$ where now the coordinate system is non-Cartesian
and has a Jacobian associated with it.

The general form of $\tilde{\rho}$ is determined by minimising the energy of the
deformed system, subject to particle conservation. For linear elastic theory, we
only need to consider only a perturbative strain on the system and expect that,
since the strain is perturbative, the density change $\tilde{\rho}$ must be
analytic in the strain tensor. We truncate the Taylor expansion to
$\bigO(\delta^2)$ in the strain tensor and use the ansatz
\begin{equation}\label{EQ:rho tilde expansion strain}
    \tilde{\rho} (\vec{r}) = \rho_1^{ik} (\vec{r}) \, \epsilon_{ik} + \rho_2^{iklm} (\vec{r}) \, u_{ik} u_{lm} 
\end{equation}
which we will justify later. This allows us to expand the new normalisation
condition for particle conservation.
\begin{equation}\label{EQ:strained normalisation condition}
    \int_\Omega (\rho_0 + \tilde{\rho}) \left( 1 + \epsilon_{ll} + M_{iklm} \, u_{ik} u_{lm} \right) \dif \vec{r} = \int_\Omega \rho_0 \dif \vec{r} 
\end{equation}
and after collecting powers of the strain tensor obtain normalisation
requirements for the different terms in the expansion
\begin{equation}\label{EQ:rho tilde normalisation conditions}
    \begin{split}
        \int_\Omega \left( \rho_1^{ik} + \rho_0 \delta^{ik} \right)\dif \vec{r} &= 0 \\
        \int_\Omega \left[
        \rho_2^{iklm} + \rho_0 \left( M_{iklm} - \delta_{ik} \delta_{lm} \right)\right] \dif \vec{r} &= 0 \, .
    \end{split}
\end{equation}

Due to the changing energy of the system (and indeed the change in average
density), we can expect to see a change in the chemical potential $\mu$ too. In
a similar spirit as that of equation \eqref{EQ:rho tilde expansion strain}, we
can analytically expand the new chemical potential in the strain tensor as 
\begin{equation}\label{EQ:mu expansion strain}
    \mu = \mu_0 + \mu_1^{ik} \, \epsilon_{ik} + \mu_2^{iklm} \, u_{ik} u_{lm} \quad .
\end{equation}
where the first term is the chemical potential of the undeformed system, and the
constants $\mu_1^{ik}$ and $\mu_2^{iklm}$ encodes how the chemical potential
changes due to the deformation and are to be determined by the least-action
principle.

We additionally need to consider how the displacement field $u_{ik}$ changes the
interaction potential $U$.
We note that the interaction is a function of separation between particles, but
importantly is originally written in terms of the separation $\Delta r'$ as
measured in the \textit{real lab frame coordinates}: $U = U(\Delta r')$ . 
For a uniform strain, we find immediately that the lab-frame separation $\Delta
 \vec{r}'$ can be written in terms of the material-frame separation $\Delta
 \vec{r}$ such that
\begin{equation}
    \left( \Delta r' \right)^2 = \left( \delta_{ik} + 2 \epsilon_{ik} + 2 \Delta_{ik} \right) \left( \Delta r \right)_i \left( \Delta r \right)_k  
\end{equation}
which is exactly equivalent to that of Landau \& Lifshitz
\cite{landau1986theory}. The tensor in the above equation is sometimes referred
to as the finite strain tensor. 
A Taylor expansion of the interaction for small strain gives
\begin{equation}
\begin{split}
    U(\Delta \vec{r}') &= U(\Delta \vec{r}) + \left( \epsilon_{ik} + \Delta_{ik} \right) f_{ik}(\Delta \vec{r}) \\
    &\quad + \epsilon_{ik}\epsilon_{lm} W_{iklm}(\Delta \vec{r})
    \end{split}
\end{equation}
with 
\begin{equation}\label{EQ:f and W definitions}
    \begin{split}
        f_{ik}(\vec{r}) &\equiv \frac{r_i r_k}{|\vec{r}|} \frac{\partial U (|\vec{r}|)}{\partial |\vec{r}|} \\ 
        W_{iklm}(\vec{r}) &\equiv\frac{ r_i r_k r_l r_m}{2 |\vec{r}|^2}
        \left( \frac{\partial^2 U}{\partial |\vec{r^2}|} - \frac{1}{|\vec{r}|} \frac{\partial U}{\partial  |\vec{r}|} \right) \; .
    \end{split}
\end{equation}

To summarise, we have now defined a perturbative deformation which changes the
geometry of our system, and have obtained the volume changes and curvature
changes due to said deformation. We have also derived a new density and chemical
potential, both of which we will solve for using the least-action principle. We
now have all the ingredients to formulate a theory of linear response of the
system to elastic deformations.

\section{Elasticity Theory}\label{sec:elasticity}

We can now develop a linear elastic theory by expanding the Lagrangian to second
order in the strain tensor and finding the response, $\tilde{\rho}$, of the
density to a small applied strain using the principle of least action. The full
details are given in Appendix \ref{app:lagrangian expansion} where we calculated
and collected all the first- and second-order components of $u_{ik}$ in the
Lagrangian. This should be of the form: 
\begin{equation}\label{EQ:lagrangian analytic expansion strain}
    L(\Omega') \simeq L_0(\Omega) + |\Omega| \pi_{ik} u_{ik} - \frac{|\Omega|}{2} A_{iklm} u_{ik} u_{lm} 
\end{equation}
where the first term is the ground-state Lagrangian, the second term contains
the coupling to the stress tensor $\pi_{ik}$, and the third term is related to
the Cauchy elastic tensor $B_{iklm}$, as carefully discussed in Bavaud et al.
\cite{bavaud1986statistical}. ($|\Omega|$ is the total volume of the
pre-deformed system.)

Let us consider the term linear in the strain tensor.
The pressure tensor is related to the stress tensor $\pi_{ik}$ by $P_{ik} =
- \pi_{ik}$. 
We will be studying a two-dimensional supersolid whose spatial modulation forms
a triangular lattice. The $C_3$ symmetry of the ground state dictates that $P_{ik} = P\delta_{ik}$, i.e.~the pressure is isotropic.
Collecting the $\bigO(u_{ik})$ terms in the expansion of the Lagrangian [see \eqref{linear terms simplified} and \eqref{pressure equation}], we can identify the
pressure as 
\begin{equation}
    \begin{split}
    P_{ik} = \int_\Omega &\left[ \frac{\hbar^2}{4m}\left(
     \frac{\partial_i \rho_0 \partial_k \rho_0}{\rho_0} - \nabla^2\rho_0 \,\delta_{ik}\right)\right.\\
     & \left. \qquad\qquad\qquad - \frac{1}{2}\rho_0 (f_{ik} * \rho_0)  \right]\frac{\dif \vec{r}}{|\Omega|}
     \end{split}
\end{equation}
which will be used in subsequent calculations. The above expression can be
directly calculated for any ground state density $\rho_0$. In the case of a
spatially homogeneous superfluid, it leads to what we expect: $P_{ik} =
\frac{1}{2}\mu_0 n \delta_{ik}$. (See Appendix A.) 

Let us now turn to the the second-order terms in the expansion
\eqref{EQ:lagrangian analytic expansion strain}. It is important to note that
the tensor $A_{iklm}$ is \textit{not} the Cauchy elastic tensor $B_{iklm}$
which provides the tensor in the Hooke's law for the stress/strain relationship
between a deformed state stress and the strain that induced the deformation,
i.e. 
$\pi_{ik} ( \Omega' ) = \pi_{ik} ( \Omega ) - B_{iklm} (\Omega) u_{lm}$.
It can be shown \cite{bavaud1986statistical} that:
\begin{equation}\label{EQ:cauchy elastic tensor definition} 
    B_{iklm} = A_{iklm} + P_{ik} \delta_{lm} - P_{im} \delta_{kl} 
\end{equation}
which we will derive independently later in \eqref{EQ:work B_iklm} by
considering work done by the expansion on the surroundings. 

For a supersolid with $C_3$ symmetry, we also expect the elastic tensor to obey
the indicial symmetries $B_{iklm} = B_{klim} = B_{lkmi} = B_{mlki}$ so that it
only contains two independent quantities --- the bulk modulus $K = c_{44}$ and
the shear modulus $G = c_{66}$: 
\begin{equation}\label{EQ:B_iklm in terms of K and G}
    B_{iklm} = K \delta_{ik} \delta_{lm} + G \left( \delta_{il} \delta_{km} + \delta_{im} \delta_{kl} \right) \, . 
\end{equation}
It should be noted that $A_{iklm}$ is not indicially symmetric in the same way
as $B_{iklm}$. Elastic theory \cite{landau1986theory} in 2D posits that the
bulk modulus $K$ is given by $B_{xxyy}$, and the shear modulus $G$ is given by
$B_{xyxy}$ (or all equivalent indicially symmetric elements). 
For a triangular lattice or a homogeneous system, the tensor also possesses the
symmetry: 
$B_{xxxx} = B_{xxyy} + 2B_{xyxy}$. Note also that we expect
the changes to the chemical potential to have the same symmetry as the pressure,
i.e. $\mu_1^{ik} = \mu_1 \delta_{ik}$. We will later make use of these
properties to simplify the expressions for the elastic constants.

We can deduce the elastic tensor by identifying the
$\bigO(u_{ik}^2)$ terms in the expansion of the Lagrangian with $A_{iklm}$, and
then using \eqref{EQ:cauchy elastic tensor definition} to obtain $B_{iklm}$. (See Appendix (\ref{app:quadratic lagrangian
density}) and the derivative rules in Appendix (\ref{app:derivative tensors})). We find that
 $B_{iklm} =\abs{\Omega}^{-1} \int_{\Omega} b_{iklm} (\vec{r}) \dif \vec{r} $ where
\begin{widetext}
\begin{equation}\label{EQ:Cauchy elastic tensor} 
        \begin{split}
            b_{iklm}(\vec{r}) =&
            + \frac{\hbar^2}{4m} \left( \frac{\partial_i \rho_0 \partial_l \rho_0}{\rho_0} \delta_{km} 
            + \frac{\partial_k \rho_0 \partial_l \rho_0}{\rho_0} \delta_{im} 
            - \frac{\partial_i \rho_0 \partial_k \rho_0}{\rho_0} \delta_{lm}  \right) \\ 
            & 
            + \frac{1}{2} \left( f_{il} * \rho_0 \right) \rho_0 \delta_{km} 
            - \frac{1}{2} \left( f_{ik} * \rho_0 \right)\rho_0 \delta_{lm} + \frac{1}{2} \left( f_{im} * \rho_0 \right) \rho_0 \delta_{kl}\\
            &
            + \left[ \left( \left( U \delta_{ik} + 2 f_{ik} \right) \delta_{lm} 
            + W_{iklm} \right)*\rho_0 \right]\rho_0 
            + \rho_1^{lm} \left[ 
            \frac{\hbar^2}{4m} \left( 2 \frac{\partial_{ik} \rho_0}{\rho_0} - \frac{\partial_i \rho_0 \partial_k \rho_0}{\rho_0^2} \right)
            + \left( f_{ik} + \delta_{ik} U \right) * \rho_0 + \mu_1^{ik} \right]
        \end{split}
    \end{equation}
    with $(g * \rho)_{\vec{r}} \equiv \int \dif \vec{r}' g(\vec{r}-\vec{r}')
    \rho(\vec{r}')$ for any function $g$ and $\rho$ and where $f_{ik}$ and
    $W_{iklm}$ are defined in \eqref{EQ:f and W definitions}. 

We note that this result for the elastic constants depends on the ground state
density $\rho_0$ as well as the shift in the density $\rho_1$ and shift in the
chemical potential $\mu_1$ as defined in \eqref{EQ:rho tilde expansion strain}
and \eqref{EQ:mu expansion strain}. In other words, we need to deduce how the
density profile has changed in response to the applied strain. This is achieved
by solving the Euler-Lagrange equations for the variable $\tilde{\rho}$ as
defined in \eqref{EQ:density analytic expression}. This involves solving the
following equation
    \begin{equation}\label{EQ:full EoM in rho_tilde}
        \begin{split}
            &\frac{\hbar^2}{4m} \grad \cdot \left( \frac{\grad \tilde{\rho}}{\rho_0} \right) - U * \tilde{\rho} \\[5pt]
            + &\frac{\hbar^2}{4m} \left( \frac{(\nabla \rho_0)^2}{\rho_0^3} - \frac{\nabla^2 \rho_0}{\rho_0^2} \right) \tilde{\rho}
            \end{split}
            \text{\Large$ = $} 
            \begin{split}
            &- \left( \mu_2^{iklm} + \mu_1^{ik} \delta_{lm} + \mu_0 M_{iklm} \right) u_{ik} u_{lm} 
            + \epsilon_{ik} \left[ -  \mu_1^{ik} + \frac{\hbar^2}{4m} \left( 2 \frac{\partial_{ik} \rho_0}{\rho_0}
            - \frac{\partial_i \rho_0 \partial_k \rho_0}{\rho_0^2} \right) \right. \\[5pt]
            & \left. + \int_\Omega \left( f_{ik}(\vec{r} - \vec{r}') + \delta_{ik} U (\vec{r} - \vec{r}') \right) \rho_0 (\vec{r}') \dif \vec{r}'
            \right]
        \end{split}
    \end{equation}
\end{widetext}
which is an integro-differential equation that can be solved either by using
Bloch's theorem (only applicable in periodic boundary conditions) or solving
directly via writing the linear operator as a matrix and using standard
iterative matrix solvers. More details on solving the equation and finding the
chemical potentials are provided in Appendix \eqref{app:full EoM in rho_tilde}. 

It must be noted that we actually have 8 equations to solve, as for each pair of
indices $(i,k)$ there is a different RHS source function, and for each of these
we need to solve the appropriate equation for both $\rho_1^{ik}$ and
$\mu_1^{ik}$. This generates 8 equations for a 2D system and 18 equations for a
3D system. Once we have the solutions for both $\rho_1^{ik}$ and $\mu_1^{ik}$,
we can calculate the other functions in the expression for \eqref{EQ:Cauchy
elastic tensor} and obtain the elastic tensor $B_{iklm}$. 

Let us check these results for the case of a homogeneous system, where the
modulation vanishes and we can check our results against the known results for
elastic moduli of a superfluid. We can see that the elastic tensor is greatly
simplified by the disappearance of the derivative terms $\partial_i \rho_0$, and
the chemical potential is analytically known: $\mu = U * \rho_0$. We can find
the convolutions $f_{ik} * \rho_0$ and $W_{iklm} * \rho_0$, solve the equation
(\ref{app:full EoM in rho_tilde}) and use this to find that 
\begin{equation}
    B_{iklm} = \mu n \delta_{ik} \delta_{lm} \quad . 
\end{equation}
Using the definitions of $K$ and $G$ from the Cauchy elastic tensors
\eqref{EQ:B_iklm in terms of K and G}, we see here that
we obtain the expected results of $K = \mu n$ and no shear modulus: $G=0$.

We will now present the numerical calculation of these homogenisation results
for the elastic constants. The results are shown in \figref{FIG:elastic constant
comparison} (solid markers). 
In our numerics, we see some small asymmetry, e.g.~$B_{iklm} \approxeq
B_{lmik}$, in the results for the elastic tensor. The discrepancy with the ideal
form \eqref{EQ:B_iklm in terms of K and G} decreases as we reduce the
discretisation and increase the system size in the numerics. With that in mind,
we choose to define the bulk modulus $K$ as the average of the values of
$B_{xxyy}$ and $B_{yyxx}$, the pressure $P$ as the average of $P_{xx}$ and
$P_{yy}$, and the shear modulus $G$ as the average of $B_{xyxy}$, $B_{xyyx}$,
$B_{yxyx}$ and $B_{yxxy}$. We then use the minimal and maximal values that the
elastic moduli \textit{could have taken} in order to generate `error bars' for
the elastic constants. It should be stressed that these error bars are not
measures of some statistical error, but rather a quantification of how much the
elastic constants deviate due to the crystal symmetry being only approximately
triangular (as opposed to exactly triangular) as a result of discretisation
error in our numerics. 

\begin{figure}
    \includegraphics[width= \columnwidth, height=0.4\textheight]{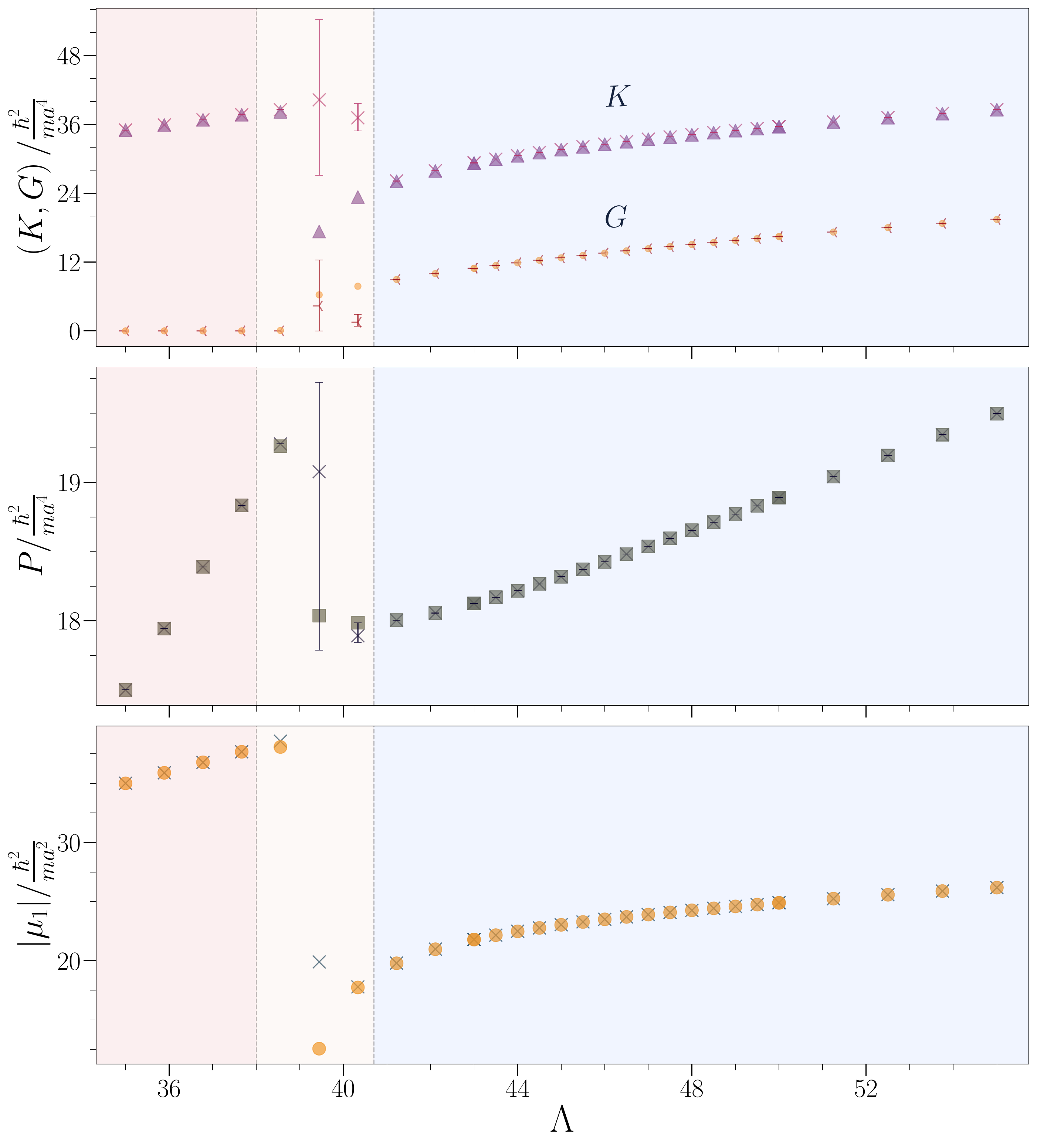}
    \caption{A comparison of the various elastic constants obtained via
    homogenisation to those from numerics (homogenisation constants are shown in
    solid markers, numerics in crosses). The top figure shows the bulk modulus
    and shear modulus, the middle figure shows the pressure, and the bottom
    figure shows the absolute value of $\mu_1$. The agreement between
    homogenisation and numerics is very strong on the whole, but falls apart in
    the phase transition region. This is likely due to numerical instabilities
    close to the phase transition which affect the numerically obtained
    constants but not the homogenisation constants.}\label{FIG:elastic constant
    comparison}
\end{figure}

We now turn to comparing these numerical results from homogenisation with a
direct calculation of the elastic constants by finding how the energy of the
ground state of the Gross-Pitaevskii equation changes when we deform the
simulation cell at fixed particle number (with periodic boundary conditions). We
begin with an undeformed ground-state obtained via numerically solving the
Gross-Pitaevskii with some given average density, area and interparticle
interaction. We then apply a small strain to the system by changing the geometry
of the simulation cell by a strain tensor $u_{ik}$, whilst keeping the total
number of particles fixed, and solving the Gross-Pitaevskii equation for the new
geometry to find the deformed ground-state. For sufficiently small strain
$u_{ik}$, we expect that such a procedure will yield a ground state with a
Lagrangian of the form \eqref{EQ:lagrangian analytic expansion strain}, where
the tensor $u_{ik}$ is now a control parameter in our computational experiment.
This strain is small in the same sense as defined in the
expansion of the Lagrangian. More precisely, we consider $u_{ik} \ll 1$, and
then try to find the harmonic terms in the energy perturbation.

By numerically extracting the Lagrangian for a system which has been deformed by
some strain $u_{ik}$, we are able to find the pressure and elastic constants by
judicious selection of strains and subsequent fitting of the coefficients. The
full procedure is described in Appendix \ref{app:numerical elastics}, where we
show the extraction of elastic constants specifically for a 2D case in a
triangular geometry. 


Our results are presented in \figref{FIG:elastic constant comparison}, where we
have plotted the bulk and shear modulus $K$ and $G$, the pressure $P$ and the
shift in the chemical potential for unit strain: $\mu_1^{ik} = \mu_1
\delta_{ik}$. These elastic constants should ideally be the same as those
obtained via homogenisation theory. In reality, they suffer from numerical
errors due to being unable to perform an arbitrarily small strain, as well as
convergence errors in the new ground-state when close to the first-order phase
transition between the homogeneous superfluid and the supersolid. 
Scanning across the whole range of the interaction parameter $\Lambda$, we find
that, outside of a small region where the phase transition occurs, the elastic
constants obtained numerically are in excellent agreement with those obtained
analytically, with a typical discrepancy of $\sim0.1\%$. 

We found that an applied numerical strain which was smaller than
$\sim 5\%$ (by this we mean $u_{ik} \sim 0.05$) would generate this excellent
agreement, but if the applied strain was $\sim 10\%$, the numerical value would
diverge from that of homogenisation. We suspect that this is due to the
introduction of anharmonic elastic moduli which are now no longer negligible.


In summary, we have developed an elastic theory for a supersolid using
homogeneration theory and calculated its elastic properties using the solution
of the Gross-Pitaevskii equation. Our homogenisation results agree well with
results from the numerical simulation of a strained supersolid.

\section{$U(1)$ Phase and Fractional Inertia}\label{sec:phasegradient}

In this section, we consider the application of a phase perturbation to the
meanfield supersolid, and analyse the coupling of the phase to the strain field.
We follow the metholodology of Josserand et al \cite{josserand2007patterns}, but
our results have some key differences that we will point out. 

We begin by specifying the nature of the phase perturbation. We consider
imposing a small superflow to the system in the form of a phase gradient. This
is applied to the supersolid \textit{in the deformed state}, i.e the lab frame
of the deformed material. The system will respond to this long-wavelength
perturbation by generating phase fluctuations, $\tilde{\phi}$, at small
wavelengths within a unit cell of the supersolid. As with $\tilde{\rho}$ in the
previous section, $\tilde{\phi}$ is a correcting function to the respective
perturbation applied to the ground state. Namely, $\tilde{\rho}$ is the function
which allows the density to settle into the energetically optimal configuration
for a given strain, and $\tilde{\phi}$ is the function which will allow the
phase field to settle into the energetically optimal configuration for a given
applied phase gradient. We mathematically represent this as 
\begin{equation}
    \phi(\mathbf{r}') = \phi_0(\mathbf{r}') + \tilde{\phi}(\mathbf{r}) 
\end{equation}
where $\phi(\mathbf{r}')$ is the total phase in the lab frame $\mathbf{r}'$,
$\phi_0(\mathbf{r}')$ is the applied perturbative phase in the lab frame, and
$\tilde{\phi}(\mathbf{r})$ is the phase-correction provided by the material in
the material frame of reference. We express $\tilde{\phi}$ in terms of the
material frame deformed coordinate system $\mathbf{r}$ in keeping with linear
response theory, similar to how we expressed $\tilde{\rho}$ in terms of the
material frame coordinate system $\mathbf{r}$. 

In order to apply perturbation theory to the phase field, we specify that
$\partial_i \phi_0(\mathbf{r}')$ is small and constant within the scope of the
unit cell of the supersolid, in a similar way to the strain tensor $u_{ik}$.
This is equivalent to stating that the phase change across the unit cell is much
smaller than $2 \pi$, i.e. $\Delta \phi \ll 2 \pi$. Note also that we are not
considering the generation of vortices.

We note that since the phase is applied in the deformed frame, the total phase
in the material frame can be expanded to obtain 
\begin{equation}\label{EQ:phi definition Taylor expansion}
    \phi(\mathbf{r}') = \phi_0(\mathbf{r}) - \mathbf{u} \cdot \grad \phi_0 (\mathbf{r}) 
    + \tilde{\phi} (\mathbf{r}) 
\end{equation}
where $\vec{u}$ is the displacement as defined in \eqref{EQ:deformation}. This
is the final expression for the phase field in the material frame of reference.
We can now proceed to expand the Lagrangian in terms of the phase field, using
the least action principle to find $\tilde{\phi}$ and the associated energy. 

Using the transformation rules for the Lagrangian (Jacobian, derivatives etc.)
and the $\phi$ expansion given by \eqref{EQ:phi definition Taylor expansion},
the phase-dependent part of the Lagrangian \eqref{EQ:GP Lagrangian Madelung} can
be expanded to $\bigO(\tilde{\phi}^2)$ as 
\begin{equation}\label{EQ:L_phi with A WITH strain}
    \begin{split}
        \mathcal{L}_{\phi} =& - \int_\Omega \Big\{\hbar \rho_0 \partial_t \phi_0 \left(1+\grad \cdot \vec{u} \right)\\ 
        &  + \frac{\hbar^2\rho_0}{2m} \left[ \left( \grad \phi_0 \right)^2 + 2 \vec{A} \cdot \grad \tilde{\phi} 
        + ( \grad \tilde{\phi})^2 \right]\Big\} \dif \vec{r} 
    \end{split}
\end{equation}
where
\begin{equation}\label{EQ:phi tilde and A ansatz WITH strain}
    \vec{A} = \grad \phi_0 + \left( \grad \phi_0 \cdot \grad \right) \vec{u} + \frac{m}{\hbar} \partial_t \vec{u}\,.
\end{equation}

The vector field $\vec{A}$ is a kind of convective flow. First of all, recall
that the superfluid velocity is given by $\vec{v} = (\hbar / m) \grad{\phi_0}$.
Also, we can define the material derivative of a field as
\begin{equation}
    \frac{D}{Dt} = \partial_t + \vec{v} \cdot \grad 
\end{equation}
to describe the time derivative of a function which is co-moving with the
superfluid velocity field. 

Note here the appearance of the term $m/\hbar \partial_t \vec{u}$ which we
previously stated as `small' without further elaboration. Here we define this
term to be small on the same order as $\vec{\nabla}\phi_0$, i.e. that the
velocity induced by the elastic deformation is comparable to the velocity
induced by the phase change.

We can rewrite $\vec{A}$ in term of these variables: 
\begin{equation}
    \vec{A} = \frac{m}{\hbar} \left( \vec{v} + \vec{v} \cdot \grad \vec{u} + \partial_t \vec{u} \right) = \frac{m}{\hbar}\left(\vec{v} +  \frac{D \vec{u}}{D t} \right)\,.
\end{equation}
A point of subtlety is that the actual deformation we have applied is $\vec{r}'
= \vec{r} - \vec{u}$, and so  
$(\hbar/m)\vec{A} = \vec{v} -D(-\vec{u})/Dt$ is a relative velocity between the
superflow and the motion of the lattice deformations.  This is exactly the form
predicted by Son \cite{son2005effective} from symmetry based arguments. In fact,
all terms in $\mathcal{L}_\phi$ satisfy the Galilean symmetry requirements
imposed by Son. In this way we can show that we have microscopically satisfied
the necessary Galilean symmetry requirements imposed on any supersolid system
macroscopically. 

Our phase Lagrangian $\mathcal{L}_{\phi}$ in \eqref{EQ:L_phi with A WITH strain}
differs from the analogous form in Josserand et al \cite{josserand2007patterns}
by the existence of the term $\hbar \rho_0 \partial_t \phi_0 \grad \cdot
\vec{u}$. This term is in fact necessary to ensure that $\mathcal{L}_{\phi}$
satisfies Galilean symmetry constraints as pointed out by Son
\cite{son2005effective}. We also note that this affects the velocity of the
excitations of the system, specifically in reference to \eqref{EQ: integrated
out Lagrangian} where this term is directly responsible  for the $n$ component o
$\hbar \left( n - \varrho + \mu_1 / \mathcal{E}'' 
\right) \dot{\phi} \vec{\nabla} \cdot \vec{u}$. If one we re to calculate th
velocities without this critical term, one would  see that it differs
significantly from those predicted by Bogoliubov  theory (shown in Section
\ref{sec:bogoliubov}).

\begin{figure}[htb]
    \includegraphics[width=\columnwidth]{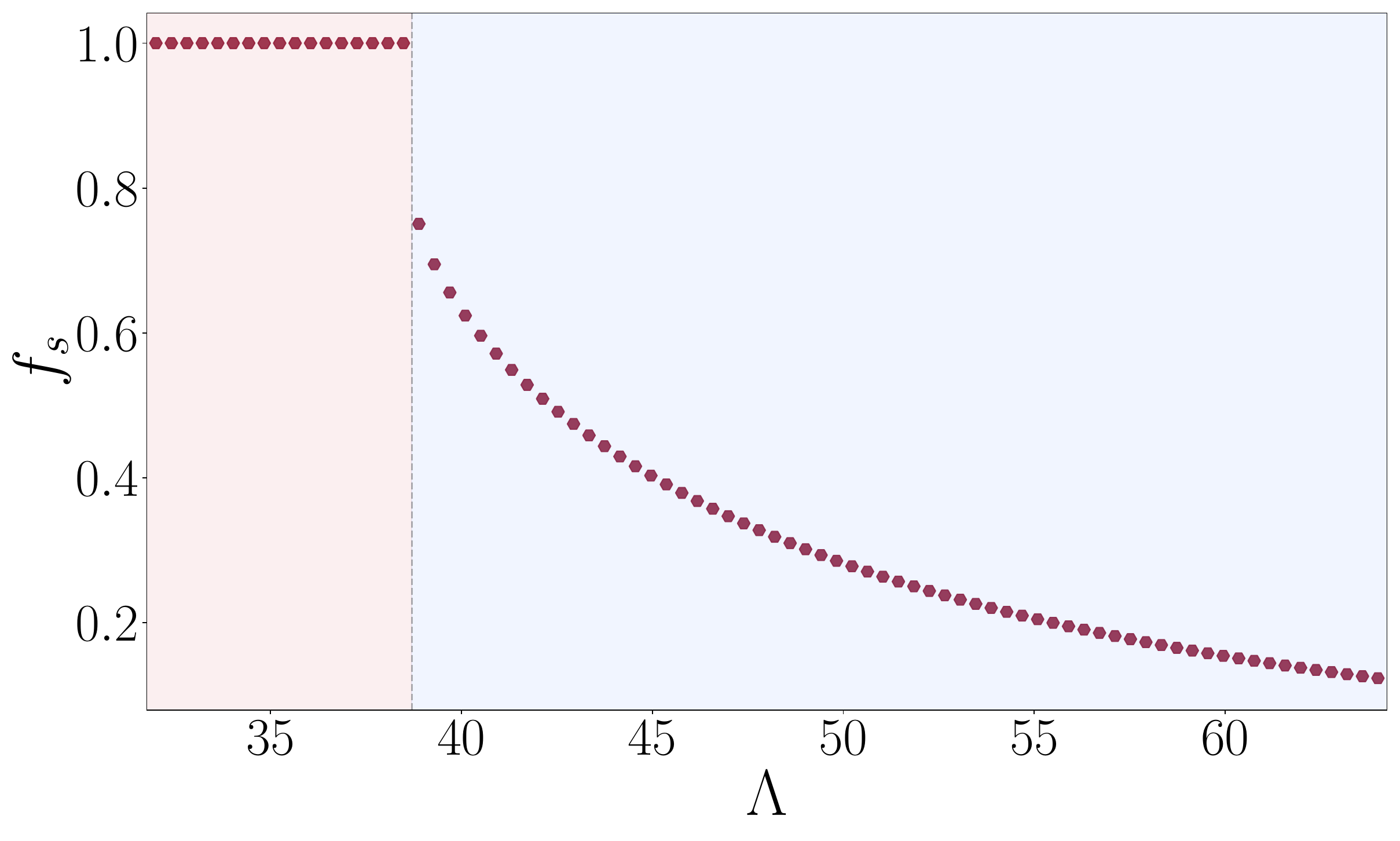}
    \caption{The superfluid fraction $f_s \equiv 1 - \varrho / n$
     as a function of $\Lambda$. We appear to observe a discontinuous transition
     at $\Lambda \sim 38$, which remains even after checking the transition
     region with a much finer $\Lambda$ discretisation. This appears to be
     numerical verification of \cite{during2011theory}, which predicted a first
     order transition close to this value.}
    \label{FIG:supersolid fraction} 
\end{figure}

Let us now solve for the short-distance phase response $\tilde{\phi}$ within a unit cell due to superflow and strain. We follow a similar procedure to Ref \cite{josserand2007patterns}. By recognising that the externally imposed phase twist and strains $\grad\phi_0$ and $\partial_i u_k$ are defined as long-wavelength relative to the unit cell, we can treat $\vec{A}$ as constant within a unit cell. However, a uniform velocity with non-uniform density does not necessarily obey local mass conservation in equilibrium. So we need a correction to the phase field, $\tilde{\phi}$, to ensure that our system obeys the continuity equation in the ground state: 
$\grad \cdot \left( \rho \grad \phi \right) = 0$. 
The Euler-Lagrange equation for $\tilde{\phi}$ gives
\begin{equation}
    \tilde{\phi} = K_i A_i\quad\mbox{with}\quad\partial_i \rho + \vec{\grad} \cdot \left( \rho \vec{\grad} K_i \right) = 0 \,.
\end{equation}
This can be solved numerically in much the same way as \eqref{rho_1 equation of
motion}. Physically, this result expresses that the fact the phase twist (and
therefore superflow) is not uniform within a unit cell because it costs less
kinetic energy to introduce phase shifts in regions of low density compared to
regions of high density. Using this solution for $\tilde{\phi}$, it can be shown that the phase-dependent
Lagrangian can be written as: 
\begin{equation}\label{EQ:L_phi with vr_ij and strain}
    \begin{split}
        L_\phi = &\abs{\Omega} \frac{\hbar^2}{2m} \varrho_{ij} A_i A_j 
        - \int_\Omega \Big[\hbar \rho_0 \partial_t \phi_0 \left( 1+ \grad\cdot\vec{u} \right)\\
       &\qquad\qquad\qquad\qquad + \frac{\hbar^2\rho_0 }{2m} \left( \grad \phi_0 \right)^2 \Big]\dif \vec{r}
    \end{split}
\end{equation}
with 
\begin{equation}\label{EQ:rho_ij definition}
    \varrho_{ij} = - \frac{1}{\abs{\Omega}} \int_\Omega \rho_0 \partial_i K_j \dif \vec{r} 
\end{equation}
which can be interpreted as a correction to the superfluid phase stiffness (see
equation \eqref{EQ:full Lagrangian with elastics and material derivatives}). Figure \ref{FIG:supersolid fraction} shows the evolution of this quantity as we increase the interaction strength $\Lambda$ showing a discontinuous transition at the superfluid-supersolid transition.

To restate the above, by carefully considering the dynamics of both an
applied/induced phase and a strain deformation (both of which are
long-wavelength relative to the unit cell), we calculate how the phase-dependent
dynamics of our system are dependent on the parameters $n$ and $\varrho_{ij}$
given by the form \eqref{EQ:L_phi with vr_ij and strain}. $n$ is simply the
density of particles in the system, and corresponds to the phase-coherent BEC
with components at both $\vec{k}=0$ and $\vec{k} \neq 0$. We will later show
that $\varrho_{ij}$ is the part of the system that corresponds to the pattern
formation, i.e. it only exists when $\psi_{\vec{k} \neq 0} \neq 0$.
Consequently, we refer to $\varrho_{ij}$ as being the ``supersolid fraction" of
the system. In this way we have separated out the pattern response of the system
to the coupled phase-strain dynamical field from the standard superfluid
response.

Finally, due to the $C_3$ symmetry, second-rank tensors representing bulk
properties should be isotropic. We therefore make the assumption that
$\varrho_{ij} = \varrho \delta_{ij}$, though in principle this is not necessary.
We also note that the dependency in \eqref{EQ:L_phi with vr_ij and strain} on
$\phi_0$ and $\vec{u}$ can be taken out of the integral (due to the same
argument that $\vec{A}$ is a constant in the unit cell). 
Combining these two simplifications, we find that we can write 
\begin{equation}
    \begin{split}
        \frac{L_\phi}{|\Omega|} 
        = &- \hbar n \dot{\phi}_0 (1+ \grad \cdot \vec{u})\\
    &        -\frac{\hbar^2}{2m}\left[ n \left(\grad \phi_0 \right)^2 - \varrho \left( \grad \phi_0 + \frac{m}{\hbar} \frac{D \vec{u}}{D t} \right)^2 \right] 
    \end{split}
\end{equation}
where $n = \int_\Omega \rho_0 \dif\vec{r}/|\Omega|$ is the average particle
density. We use this result, along with the elastic expansion \eqref{EQ:Cauchy
elastic tensor}, to write out the full  Lagrangian after the
homogenisation process as 
\begin{widetext}
    \begin{equation}\label{EQ:full Lagrangian with elastics and material derivatives}
        \frac{\Lhomo}{\abs{\Omega}} = L_0 - \hbar n \dot{\phi}_0 (1+ \grad \cdot \vec{u}) - P_{ik} u_{ik} - \frac{1}{2} A_{iklm} u_{ik} u_{lm} - \frac{\hbar^2}{2m} \left( n - \varrho \right) \left( \grad \phi_0 \right)^2 
        + \hbar \varrho \grad \phi_0 \cdot \frac{D \vec{u}}{D t} + \frac{1}{2} m \varrho \left( \frac{D \vec{u}}{D t} \right)^2 
    \end{equation}
\end{widetext}  
This effective Lagrangian contains all the linear response of the system due to
both strain and phase gradients. We can see that the phase stiffness is now
given by $n - \varrho$ and $\varrho$ is the inertia associated with the motion
of the modulated pattern of the density. In summary, we have followed
Ref.~\cite{josserand2007patterns} and reproduced the same results for the
superfluid phase stiffness of the supersolid. Our Lagrangian differs from that
reference in the dynamical term. In the next section, we complete the
development of an effective low-energy theory for the supersolid by considering
the free energy of the system in the grand canonical ensemble. Then, we can
finally proceed to discuss the macroscopic dynamics of the system.

\section{Free energy and Grand Canonical ensemble} 

In Sections \ref{sec:homogenisation} to \ref{sec:phasegradient}, we developed a homogenised theory of the supersolid in the canonical ensemble at a fixed average particle density of $n$. We have integrated out the
intra-unit-cell dynamics of the supersolid and have now obtained an effective
Lagrangian $\Lhomo(\phi_0,\vec{u})$ which describes the system under a given phase gradient and displacements $\vec{u}$. We have assumed that all the unit cells responded in the same way, i.e. the response has the same periodicity as the supersolid. 
This effective Lagrangian obeys all the
necessary requirements of Galilean invariance, rotational invariance and $U(1)$
invariance as specified by Son \cite{son2005effective} and Andreev \& Lifshitz \cite{andreev1969quantum}, so we would expect the
dynamics to be phenomenologically correct. 


In this section, we will consider the inter-cell dynamics and allow for particle
flow between unit cells leading to variations in the density $\delta n$ across
the system at length scales much larger than the size of the unit cells of the
supersolid.



It is important to note that the additional energy response due to flow between
cells is crucial to obtaining the correct dynamics of the system. If we were to
neglect this energy response, we would find that the dynamics of the system
would not match up with a Bogoliubov treatment. A similar phenomena was observed
in \cite{sutton1988tight}, where the authors found that the bulk modulus
computed via a homogeneous dilation technique was not in agreement with
long-wavelength phonon predictions unless charge redistribution was taken into
account. 

We assume that the inter-cell number fluctuation is small, i.e. that $\delta n /
n \ll 1$, and so we can expand the Lagrangian analytically in $\delta n$ up to
$\bigO(\delta n^2)$. 
We now consider the ground state Lagrangian density $\mathcal{E}(n) = -
L_0/|\Omega|$ which is a function of the average density $n$ of the undeformed
system. Under a change of the local density, this can be expanded as 
\begin{equation}
    \mathcal{E}(n + \delta n) \simeq \mathcal{E}(n) + \mathcal{E}'\delta n 
    + \frac{1}{2}\mathcal{E}'' (\delta n )^2
\end{equation}
and similarly the local pressure can be expanded as 
\begin{equation}
    P(n + \delta n) \simeq P(n) + P' \delta n \, .
\end{equation}
These are the only terms we need to track in a theory that is an expansion up to
second order in the total power of all perturbative fields.
The coefficients, $P'$ and $ \mathcal{E}''$,
can be calculated numerically by solving
Gross-Pitaevskii equation at several densities. 
We can also use $\mathcal{E}' = \mu_0$ and $\mathcal{E}'' = \partial\mu/\partial n$ if we have the density dependence of the chemical potential. In fact, we have already calculated the pressure derivative in terms of the chemical potential shift per unit strain, $\mu_1$, as defined in \eqref{EQ:mu expansion strain} with $\mu_1^{ik} = \mu_1\delta_{ik}$.
To see this, we note that the pressure, $P = -(\partial E/\partial V)_N$, and
the chemical potential, $\mu = (\partial E/\partial N)_V$, are related by a
Maxwell relation so that
\begin{equation}
P' = \frac{\partial P}{\partial n} 
= V\left(\frac{\partial P}{\partial N}\right)_V \!\!\!
= -V\left(\frac{\partial \mu}{\partial V}\right)_N \!\!\! 
= -\mu_1\,.
\end{equation}
since $\delta V/V = u_{kk}$ is the strain. Similarly, we can use $\mathcal{E}''
= \partial\mu/\partial n$ if we have already obtained the density dependence of
the chemical potential.

In general, the above analysis is valid for any tensorial
quantity, but we have elected to use the symmetry properties of tensors that
belong to a triangular lattice (i.e. $P_{ik} = P \delta_{ik}$ and $\mu_1^{ik} =
\mu_1 \delta_{ik}$) to simplify our calculations. The following analysis also
uses this symmetry, but can be generalised to any tensorial quantity.


Finally, we need to include the work done by the change in local density on the
local environment.
Namely, if a part of the system expands due to an increase in particle number,
it does work on the surroundings which must compress. The work done caused by a
total volume change of $\delta V$ is:
\begin{equation}
   \delta W =  -P \delta V = -P (\epsilon_{ll} + M_{iklm} u_{ik} u_{lm}) |\Omega|\,.
\end{equation}
In order to describe the dynamics of these density functions, we need to add
this work done to the Lagrangian to obtain 
\begin{equation}
    \frac{\Ldyn}{\abs{\Omega}} =  \frac{\Lhomo}{\abs{\Omega}} -
        P \left( \epsilon_{ll} + M_{iklm} u_{ik} u_{lm} \right) \, . 
\end{equation}
Comparing with our form \eqref{EQ:full Lagrangian with elastics and material derivatives} for $\Lhomo$ and the relation \eqref{EQ:cauchy elastic tensor definition}  between the Cauchy tensor $B_{iklm}$ and $A_{iklm}$, we find that
\begin{align}\label{EQ:work B_iklm}
-& P u_{ll} - \frac{A_{iklm}}{2} u_{ik} u_{lm}\! -\! P (\epsilon_{ll} + M_{iklm} u_{ik} u_{lm})\notag\\
&\simeq - \frac{1}{2}B_{iklm}u_{ik} u_{lm} \, .
\end{align}
Using this, we arrive at an effective quadratic Lagrangian that can describe the dynamics of the low-energy excitations of the supersolid. This is the principal result of our paper: 
\begin{widetext}
    \begin{equation}\label{EQ:grand canonical lagrangian with work}
        \begin{split}
            \frac{\Ldyn}{\abs{\Omega}} = & - \mathcal{E} - \mu_0 \delta n - \frac{1}{2} \mathcal{E}'' (\delta n)^2  
            - \hbar n \dot{\phi}_0 \left(1 + \grad \cdot \vec{u} \right) - \hbar \delta n \dot{\phi}_0 + \mu_1 \delta n \grad \cdot \vec{u}  \\ 
            & - \frac{1}{2} B_{iklm} u_{ik} u_{lm} - \frac{\hbar^2}{2m} \left( n - \varrho \right) \left( \grad \phi_0 \right)^2 
            + \hbar \varrho \grad \phi_0 \cdot \dot{\vec{u}} + \frac{1}{2} m \varrho \dot{\vec{u}}^2 \,.
        \end{split}
    \end{equation}
\end{widetext}
In this process we note that we  have directly identified the Cauchy elastic
tensor in two different ways, firstly by using the result from Bavaud
\cite{bavaud1986statistical}, and secondly by directly considering the work the
excitations would need to do on their environment. 

To summarise, we have a coarse-grained Lagrangian to describe the low-energy excitations of the supersolid. In the next section, we will investigate these dynamics.

\section{Effective supersolid Lagrangian}

In the previous sections, we have derived an effective Lagrangian \eqref{EQ:grand
canonical lagrangian with work} by integrating out the short-wavelength modes in
the unit cell and coarse-graining the system such that a unit cell is now
considered a point in the continuous field theory described by the fields $\{
\vec{u}, n, \phi \}$. This Lagrangian formally describes a low-energy effective
field theory wherein the coupled Goldstone modes of the system are the low-lying
excitations. We now set out to calculate the normal modes by integrating out the
field $\delta N$ and solving the resulting set of coupled equations. 

By minimising the Lagrangian for $\delta n$, we find 
\begin{equation}
    \delta n = \frac{\mu_1}{\mathcal{E}''} \grad \cdot \vec{u} - \frac{\hbar}{\mathcal{E}''} \dot{\phi}
\end{equation}
which we substitute back into the Lagrangian to straightforwardly obtain 
\begin{equation}\label{EQ: integrated out Lagrangian}
    \begin{split}
        L = & - \mathcal{E} - \hbar n \dot{\phi} + \frac{\hbar^2}{2} 
        \left[ \frac{\dot{\phi}^2}{\mathcal{E}''} - \frac{n - \varrho}{m} \left( \grad \phi \right)^2 \right] \\
        & + \frac{1}{2} \left[ m \varrho \dot{\vec{u}}^2 - \left( B_{iklm} 
        - \frac{\mu_1}{\mathcal{E}''} \mu_1 \delta_{ik} \delta_{lm} \right) u_{ik} u_{lm} \right] \\
        & - \hbar \left( n - \varrho + \frac{\mu_1}{\mathcal{E}''} \right) \dot{\phi} \grad \cdot \vec{u} 
    \end{split}
\end{equation}
This Lagrangian should be able to describe the Goldstone modes for the system. This effective
Lagrangian is of the form predicted by Ye \cite{ye2008elementary}. 

By counting the number of spontaneously broken continuous symmetries, we are
able to predict the number of Goldstone modes that emerge. Since we have broken
the $U(1)$ gauge symmetry, we expect to find one Goldstone mode. We also have a
triangular lattice, which means we have broken two continuous translational and
rotational symmetries, and so we expect to find two Goldstone modes. Therefore
we should find three Goldstone modes in total. 

One can immediately find the equations of motion for both $\phi$ and $\vec{u}$,
but it serves as a useful check to ensure we recover the Lagrangian for a
Bogoliubov mode in the superfluid phase. In a superfluid phase, one can show
that $B_{iklm} = U_{k=0} n^2 \delta_{ik} \delta_{lm}$, $\mu_1 = - U_{k=0} n $,
$\mathcal{E}'' = U_{k=0}$ and $\varrho = 0$. Consequently, the resultant
Lagrangian is now 
\begin{equation}\label{Bogoliubov lagrangian}
    L = - \mathcal{E} - \hbar n \dot{\phi} + \frac{\hbar^2}{2} 
    \left[ \frac{\dot{\phi}^2}{U_{k=0}} - \frac{n}{m} \left( \grad \phi \right)^2 \right] 
\end{equation}
which corresponds exactly with the superfluid Bogoliubov mode.

Note that for the following derivation of the supersolid
Goldstone velocities, it is assumed that the ground state is a supersolid and
hence $\varrho \neq 0$. In a superfluid state ($\varrho = 0$), the following
analysis does not apply, and instead one needs to refer to \eqref{Bogoliubov
lagrangian}.

For the general supersolid case, we derive the equations of motion directly by
assuming the fields $\phi$ and $\vec{u}$ are of a plane wave form $\sim e^{i
\left( \vec{k} \cdot \vec{r} - \omega t \right)}$. The vector equation for the
variable $\vec{u}$ can be solved to provide both the shear and longitudinal
mode. The Euler-Lagrange equation for the variable $\vec{u}$ is immediately
found to be 
\begin{equation}\label{EQ:vector u equation}
    \begin{split}
        m \varrho \ddot{\vec{u}} - & \left( K + G 
        - \frac{\mu_1}{\mathcal{E}''} \mu_1 \right) \grad \left( \grad \cdot \vec{u} \right)  
        - G \nabla^2 \vec{u} \\ 
        & - \hbar \left(n - \varrho + \frac{\mu_1}{\mathcal{E}''} \right) \grad \dot{\phi} = 0 
    \end{split}
\end{equation}
where we have used \eqref{EQ:B_iklm in terms of K and G} as we are analysing a
triangular lattice. 

We can now solve this equation by noting that the vector field $\vec{u}$ can be
split into longitudinal and transverse elements $\mathbf{u} = \mathbf{u}_l +
\mathbf{u}_t$. Taking the curl of \eqref{EQ:vector u equation} isolates the transverse motion ($\vec{u}_t$) giving:
\begin{equation}\label{shear u equation}
    \grad \times \left( m \varrho \ddot{\vec{u}}_t - G \nabla^2 \vec{u}_t \right) = 0 
\end{equation}
This is a wave equation for $\grad\times\vec{u}$ from which we obtain the shear velocity 
\begin{equation}
    c_{\textrm{shear}} = \sqrt{\frac{G}{m \varrho}} 
\end{equation}
which is exactly equivalent to the shear velocity predicted in standard solid
elasticity. 

We examine longitudinal dynamics ($\vec{u}_l$) by taking the divergence of \eqref{EQ:vector u equation} we
are examining the dynamics of $\mathbf{u}_l$. 
We find the two coupled equations
\begin{equation}\label{EQ:phi equation of motion}
    \hbar \left( \frac{\omega^2}{\mathcal{E}''} - \frac{n - \varrho}{m} \vec{k}^2 \right) \Phi 
    - i \omega \left(n - \varrho + \frac{\mu_1}{\mathcal{E}''} \right) \grad \cdot \vec{u}_l = 0\,, \\
\end{equation}
\begin{equation}\label{EQ:vector u_l equation of motion}    
    \begin{split}
        \left[m \varrho \omega^2 - \left(K - \frac{\mu_1}{\mathcal{E}''}\mu_1 + 2 G \right) \vec{k}^2 \right]   \grad \cdot \vec{u}_l & \\
        + i \omega \hbar \vec{k}^2 \left(n - \varrho + \frac{\mu_1}{\mathcal{E}''} \right) & \Phi = 0
    \end{split}
\end{equation}
in the variables $\mathbf{u}_l$ and $\phi$. It can be shown easily that if we
are in a superfluid regime the second equation is trivially satisfied, and the
first equation finds the Bogoliubov frequency. These equations can be solved for
$\omega$ as a function of $\mathbf{k}$ and leads to the following dispersion
relation 
\begin{equation}
    \omega^2 = \frac{ A \pm D}{2 m \varrho} \, \vec{k}^2 
\end{equation}
with 
\begin{equation}
\begin{split}
        A &= K + 2 G + \left(n \mathcal{E}'' + 2 \mu_1 \right) \left( n - \varrho \right) \\ 
        D^2 &= A^2 - 4 \varrho (n-\varrho)
        \left[ \mathcal{E}''(K+2G) -\mu_1^2
        \right]\,.
    \end{split}
\end{equation}
We note that, in the case of the homogeneous superfluid ($\varrho=0$), 
we obtain the
superfluid velocity directly from \eqref{Bogoliubov lagrangian}
\begin{equation}
    \omega^2 = \frac{n \, U_{k=0}}{m} \vec{k}^2 
\end{equation}
which is exactly equal to the Bogoliubov velocity. In the opposite limit of
$\varrho=n$, the two modes have frequencies 
\begin{equation}
    \omega^2 = \frac{K+2G}{m n} \vec{k}^2 \quad\text{and zero.}
\end{equation}
The zero mode corresponds to the collapse of the Bogoliubov mode and the linear
dispersing mode has a velocity as expected for a normal solid with elastic
moduli $K$ and $G$.

It is interesting to note that, although we expect the Gross-Pitaevskii
Lagrangian to struggle with describing a normal solid, the effective
phenomenological Lagrangian that emerges actually returns to that of a normal
solid in the limit $\varrho \rightarrow n$. Moreover, although all of the modes
have a factor of $1/\varrho$ in their velocities which seems to indicate that
the velocities should blow up as $\varrho \rightarrow 0$, the velocities appear
to be well behaved as $\varrho$ becomes small. This suggests that there is some
implicit dependence on $\varrho$ hidden in the values of $B_{iklm}$ etc., which
stops the velocities from blowing up. However, we should note that as we appear
to see a first-order transition, the value of $\varrho$ never really approaches
zero in our numerics, but only jumps directly to zero at which point these
equations are no longer valid.

\begin{figure}[htb]
    \centering
    \includegraphics[width= \columnwidth]{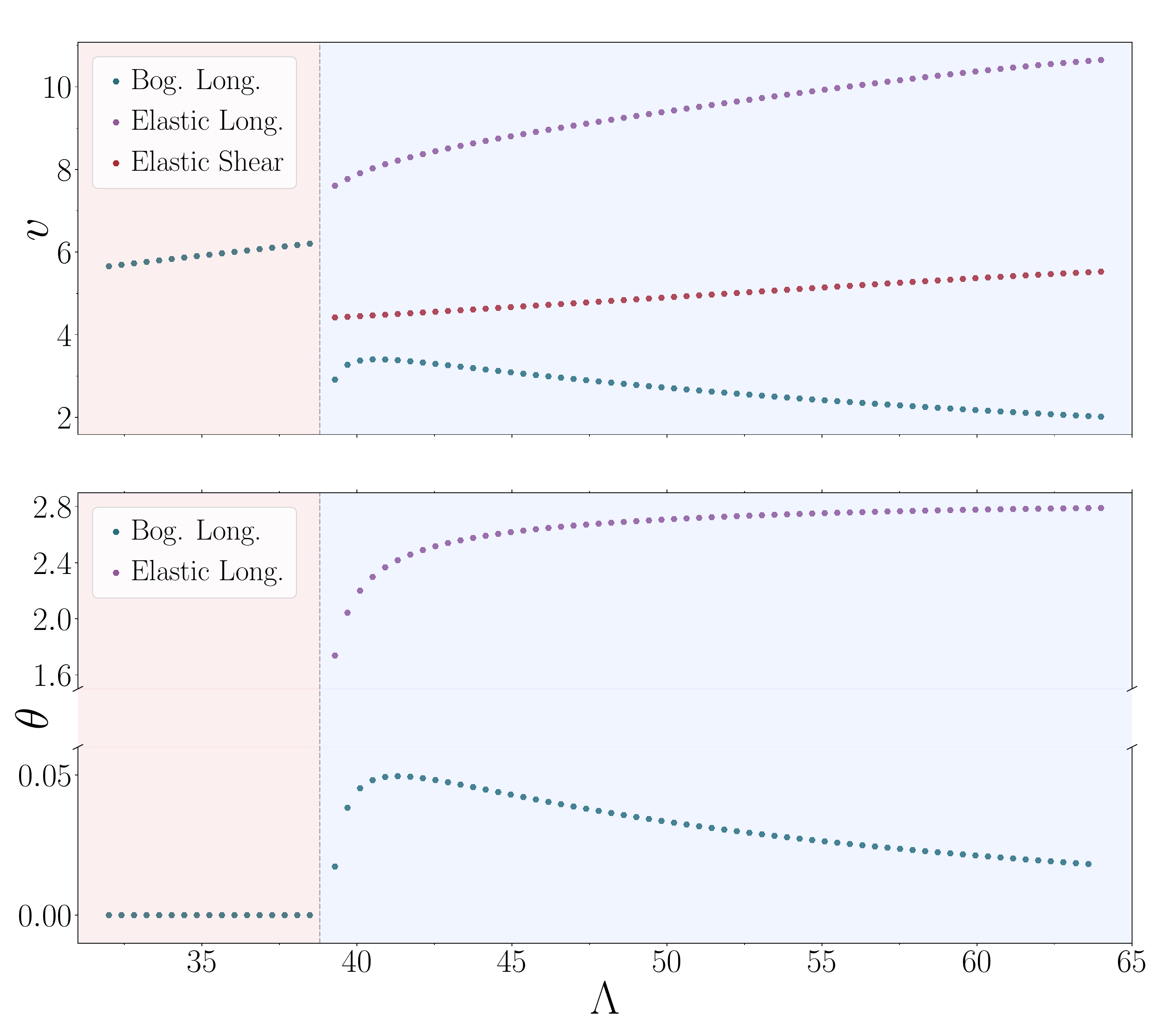}
    \caption{Top: excitation velocities (in units of $\hbar / \sqrt{n} m a^2$)
    for all available sound modes of a system governed by a Heaviside
    interaction of strength $\Lambda$ with a phase transition at $\Lambda \sim
    38$, as calculated via homogenisation. Note that in the superfluid phase the
    bulk mode is the Bogoliubov mode, hence the sudden appearance of two extra
    modes at the superfluid-supersolid transtion. Bottom: the eigenvector angle
    $\theta$ for the available longitudinal modes.}\label{FIG:sound velocities} 
\end{figure}

We can also now examine the nature of the normal modes of the system. By writing
\eqref{EQ:phi equation of motion} and \eqref{EQ:vector u_l equation of motion}
in terms of a matrix equation, we can find the eigenvectors of the system for a
given $\omega$. These eigenvectors are the normal modes, and are some composite
of the fields $\phi$ and $\mathbf{u}_{l}$. We can then subsequently define an
eigenvector angle $\theta$ which describes the relative contribution of the
fields $\phi$ and $\mathbf{u}_{l}$ to the normal mode. For example, in the
superfluid the eigenvector angle is $\theta = 0$ as the normal mode is purely
$\phi$, i.e. $\mathbf{e}_n \equiv \mathbf{e}_\phi$. A supersolid with normal
mode $\mathbf{e}_n = \left( 0.5, 0.5 \right)$ would have $\theta = \pi/4$ etc.
The normal mode composition is illustrated in \figref{FIG:sound velocities}. 

An interesting point of note is that the slow longitudinal mode (which is
expected to die off in the limit of a solid) is a composite of both $\phi$ and
$\mathbf{u}_l$, but is mostly $\phi$ at all points in the phase diagram (i.e.
$\theta$ is small). More importantly, it appears there is a critical point at
which the slow longitudinal mode is 'saturated', that is to say that there is a
global maximum in the mixing of the slow longitudinal mode (but note that the
mode is still mostly $\phi$). This feature appears to be absent in the fast
longitudinal mode. However, the fast longitudinal mode also exhibits interesting
behaviour as the mixing angle does not tend to $\pi / 2$ or $3 \pi / 2$ as one
may initially expect (as we would expect the mode to tend to a purely elastic
phonon mode in the limit of a solid), but instead tends to approximately $7\pi/8$. This is a curious feature of the fast longitudinal mode, and we are
currently investigating the origin of this behaviour.

The above holds for any system described by a Gross-Pitaevskii equation; we can
now use these results in the specific case of a soft-core interaction. We find a
ground state for the system numerically and by applying the procedure detailed
above we can determine the velocities of all modes that exist. Importantly we
note that our Lagrangian is tuned \textbf{across the transition}, i.e. a single
theory describes both the superfluid and supersolid regime and continuously
varies across a first order phase transition. Importantly, symmetry arguments by
Son \cite{son2005effective} dictate the functional form of any Lagrangian
belonging to a supersolid and since our Lagrangian is a subset of those allowed
by symmetry, we argue that we completely capture the phenomenological behaviour
of the supersolid as described by a Hartree-Fock approximation. 

\section{Bogoliubov Dispersion}\label{sec:bogoliubov} \indent An alternative
technique to finding excitations of the system is to directly apply linearised
oscillations in the number density and the phase. Namely, since $\psi =
\sqrt{\rho} e^{i \phi}$, we can apply the Bogoliubov fluctuations $\rho
\rightarrow \rho_0 + \delta \rho$ and $\phi$. We expect to recover all the
Goldstone modes accessible to the system, and expect the quantitative properties
to be the same as those predicted in the elastic theory. More precisely, due to
the broken $U(1)$ phase and the broken continuous translational symmetry, we
expect to find three Goldstone modes, two of which are longitudinal and one of
which is a shear mode. Since this is a completely independent technique to
recovering the Goldstone modes, it serves as an important check of our results. 

Whilst both techniques access the speeds of sound, the elastic technique also
provides elastic moduli and fractional inertia, whereas the bogoliubov technique
provides a full band structure. In this way, the two techniques complement each
other, as they verify common ground but also provide information that is
inaccesible to the other.

By following a similar procedure to that outlined in
\cite{kunimi2012bogoliubov}, we are able to obtain eigenvalue equations which
give the excitation frequencies $\omega$ of the Bloch waves in the system. These
Bloch waves have a band structure, and the lowest band structure at the $\Gamma$
point corresponds to the Goldstone modes of the system whose propagation
velocities were independently derived in \cite{kunimi2012bogoliubov}. The band
structure is fully described by the equations 
\begin{equation}\label{bogoliubov dispersion equations}
    \begin{split}
        & \left[ - \omega^2 + \left( \hat{T}^{\mathbf{k}} + \hat{U}^{\mathbf{k}} \right) \hat{T}^{\mathbf{k}} \right] (\Phi)^{\mathbf{k}} = 0 \\ 
        & \left[ - \omega^2 + \hat{T}^{\mathbf{k}} \left( \hat{T}^{\mathbf{k}} + \hat{U}^{\mathbf{k}} \right) \right] (\delta \psi)^{\mathbf{k}} = 0 
    \end{split}
\end{equation}
with $\hat{T}$ and $\hat{U}$ defined as
\begin{equation}
    \begin{split}
        \hat{T}^{\mathbf{k}}_{\mathbf{G}, \mathbf{G}'} = & \left[ \frac{1}{2} \left( \mathbf{G} + \mathbf{k} \right)^2
        - \mu \right] \delta_{\mathbf{G}, \mathbf{G}'} + U_{\mathbf{G} - \mathbf{G}'} \rho_{\mathbf{G} - \mathbf{G}'} \\
        \hat{U}^{\mathbf{k}}_{\mathbf{G}, \mathbf{G}'} = & 2 \sum_{\mathbf{G}''} \psi_{\mathbf{G} - \mathbf{G}''} 
        U_{\mathbf{k} + \mathbf{G}''}  \psi_{\mathbf{G}'' - \mathbf{G}'}  
    \end{split}
\end{equation}
with $U_{\mathbf{q}}$, $\psi_{\mathbf{q}}$ and $\rho_{\mathbf{q}}$ the fourier
transforms of the interaction potential, the superfluid order parameter and the
density respectively. We expect that the band structure obtained from solving
these equations should find all the requisite Goldstone modes and that they
should have the same velocities as predicted via homogenisation.

\begin{figure}[ht]
    \centering
    \includegraphics[width= \columnwidth, height=0.25\textheight]{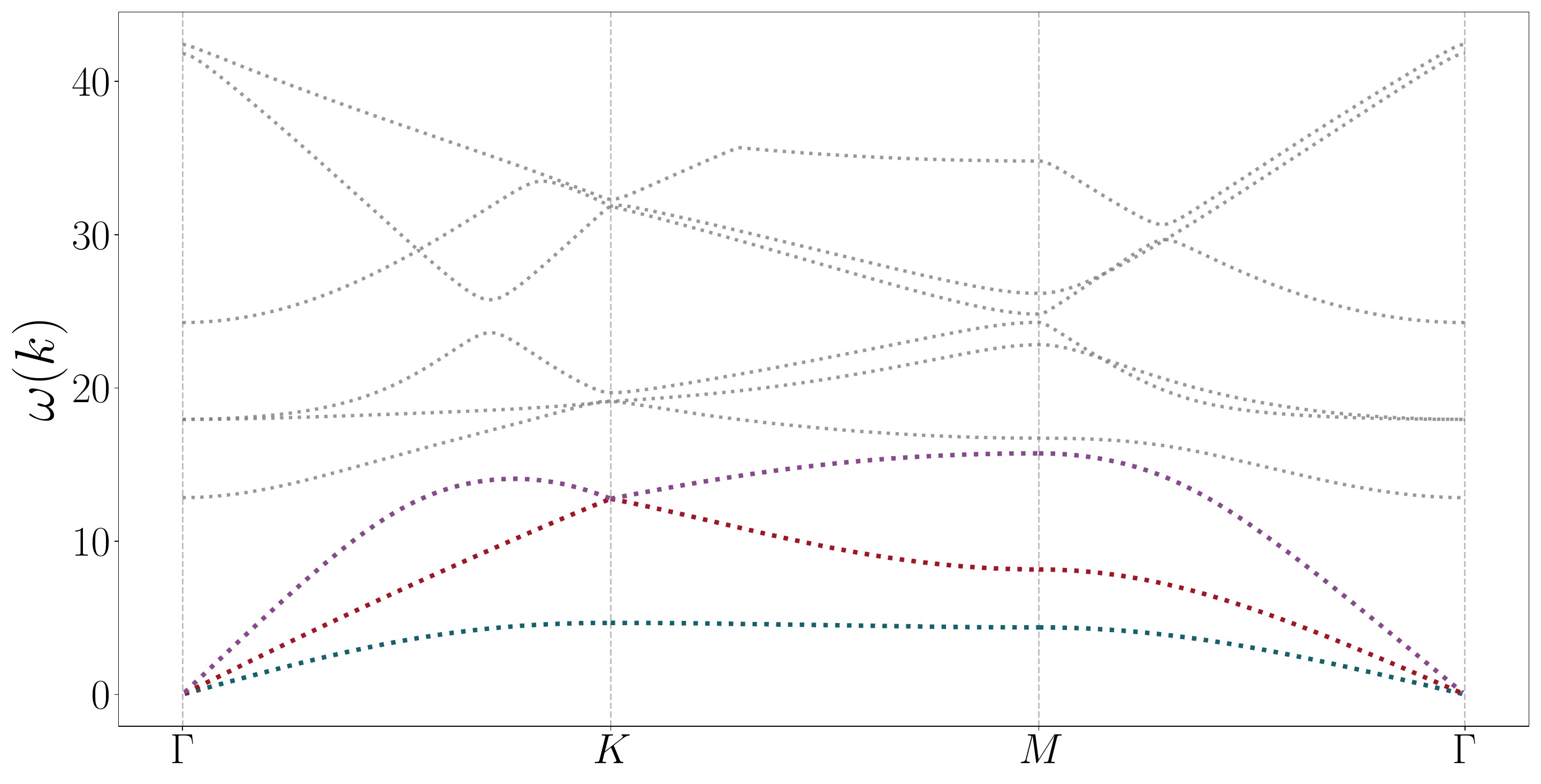}
    \caption{Band structure for a supersolid at $\Lambda = 50$. The three
    Goldstone modes are highlighted as a \textcolor{GraphGreen}{fast
    longitudinal}, \textcolor{GraphBlue}{slow longitudinal} and
    \textcolor{GraphRed}{shear} mode.} \label{FIG: Bogoliubov dispersion
    relation} 
\end{figure}

Upon recovery of the band structure, one finds exactly $(D+1)$ Goldstone
modes as expected. There is a band crossing at the $K$ point of the fast
longitudinal and shear modes, characteristic of a triangular lattice (n.b. this
is not actually a crossing but a band touching, as shown by the $M \rightarrow
\Gamma$ band structure). 

For any particular given $\Lambda$, both the Bogoliubov and elastic techniques
predict all three speeds of sound (if supersolid, only one speed if superfluid),
and agree on the speeds with excellent precision. Subsequently, by deriving the
excitations of the system in two independent ways and obtaining the same results
to excellent numerical precision, we can reliably say we have obtained the
correct dynamics. 

\section{Results and Discussion}

We can see that we have now verified the technique of homogenisation as a means
of recovering the low-lying excitations of the system. This was achieved via
means of two completely independent techniques, the results of which both agree
to excellent accuracy. An advantage of homogenisation over the Bogoliubov
technique is that it also provides the elastic constants of the system, which
cannot be obtained via Bogoliubov. Moreover, it provides the fractional inertia
of the system, which is a quantity that is of great physical interest. A clear
disadvantage however, is that homogenisation does not provide the full spectrum
of excitations, only the long-wavelength limit. We stress that this is not an
issue, as the Bogoliubov technique is completely separate from homogenisation,
and so by utilising both techniques one can recover the full band structure from
Bogoliubov and supplement it with additional information in the long-wavelength
limit from homogenisation. 

The effective Lagrangian that is the end result of homogenisation elucidates
interesting dynamics. By considering that the flow excitations of the system can
be coupled to the strain excitations, we can see that the system is capable of
strain-induced flow. Furthermore, by considering the application of an
instantaneous strain to the system, we can see that flow will be induced and
contribute to the elastic response of the system. Alternatively, if we apply a
strain and keep it fixed for a long time so that the flow dissipates, there will
be a different elastic response. In this way we can see that the elastic modulus
has a time dependence, and that the system is capable of dissipating energy via
the flow of the system. This energy dissipation is likely to be via the two
longitudinal modes which are coupled crystal and Bogoliubov phonons. 

There is an interesting question of what the compressibility of the system is.
When one considers the elastic strain, it is natural to define the
compressibility of the system as $\kappa = - 1/V \, ( \partial V / \partial P
)$. Typically in a solid, one would simultaneously be able to say $\kappa = 1 /
n \, ( \partial n / \partial P )$, where $n$ is the number density and $\mu$ the
chemical potential. However, in our analysis, we can show that these two
definitions do not agree, and are off by a considerable margin. These two values
are defined as $\kappa = 1 / B_{xxyy}$ and $\kappa = 1 / (n \mu_1 )$ in our
notation respectively. 

The resolution of this seeming dilemma is to revisit the definition of the two
compressibilities. The compressibility $\kappa = 1/B$ where $B = - V \, (
\partial P / \partial V \bigr|_N )$ is the elastic bulk modulus of the system,
taken at constant particle number. This is the compressibility of the system as
defined via elastic theory, and is the one that we will take as the 'true
compressibility', and consequently $B$ as the 'true bulk modulus'. Importantly,
the derivative is taken under constant particle number. As such, to elucidate
this a little further, we write 
\begin{equation}
    B_e = - V \frac{\partial P}{\partial V}\biggr|_{N}  
\end{equation}
where the subscript $e$ denotes that this is the elastic bulk modulus. We can
then write the compressibility as $\kappa_e = 1/B_e$. We can subsequently define
the thermodynamic bulk modulus (using the typical definition with $n$ but now
removing the factor of $V$ which is kept constant) as 
\begin{equation}
    B_t = N \frac{\partial P}{\partial N}\biggr|_{V}    
\end{equation}
and the thermodynamic compressibility accordingly. 

The key consideration is that the $B_e \neq B_t$ due to the fact that we are
keeping different variables constant upon taking derivatives. If we want to
connect these two definitions, we need to use the cyclic relations of
thermodynamics 
\begin{equation}
    \frac{\partial x}{\partial y} \biggr|_z = - \frac{\partial x}{\partial z} \biggr|_y \frac{\partial z}{\partial y} \biggr|_x \quad . 
\end{equation}
We can then write 
\begin{equation}\label{EQ:thermodynamic cyclic relation}
    B_e = - V \frac{\partial P}{\partial V}\biggr|_{N} = V \frac{\partial P}{\partial N}\biggr|_{V} \frac{\partial N}{\partial V}\biggr|_{P} 
    \equiv B_t \ \frac{1}{n} \frac{\partial N}{\partial V}\biggr|_{P} 
\end{equation} 
from which we can consider the following. In a regular material (solid, gas,
liquid) we would expect that the derivative of the particle number with respect
to the volume under constant pressure is simply the density of the system, and
we would recover the thermodynamic bulk modulus $B_t = n \, ( \partial P /
\partial n \bigr|_{V} )$ from \eqref{EQ:thermodynamic cyclic relation}. We can
then examine whether this is the case for our system, i.e. is $1 / n \, (
\partial N / \partial V \bigr|_{P} ) = 1$?

We can examine this quantity, henceforth denoted as $\mathcal{R}$ and called the
'compressibility ratio', by choosing an initial volume and particle number $V_0$
and $N_0$ which will generate a $P_0$, adjusting the volume around $V_0$ with
some $\delta V$, and then varying the particle number $N_0 + \delta N$ until the
system reaches $P_0$. We can then numerically calculate $\mathcal{R}$ by finite
differencing. If we expect the difference between the two compressibilities to
be due to the fact that $\frac{1}{n} \frac{\partial N}{\partial V}\bigr|_P \neq
1$ then we would expect $\mathcal{R} = \frac{\kappa_t}{\kappa_e}$, or
equivalently  
\begin{equation}\label{EQ: compressibility ratio definition}
    \mathcal{R} = \frac{B_e}{B_t} \equiv \frac{K}{n \mu_1} \, .
\end{equation}

Upon performing the numerical determination of $\mathcal{R}$, we find that it is
within good agreement of the ratio of the two compressibilities as determined
via homogenisation. The percentage error between the homogenisation and
numerical $\mathcal{R}$ is typically between $3\%$ and $5\%$. We believe this is
an artifact of the numerical determination of $\mathcal{R}$, and that as the
algorithm and precision of simulations for determining $\mathcal{R}$ is
improved, the agreement will improve accordingly. 

\begin{figure}[ht]
    \centering
    \includegraphics[width=\columnwidth]{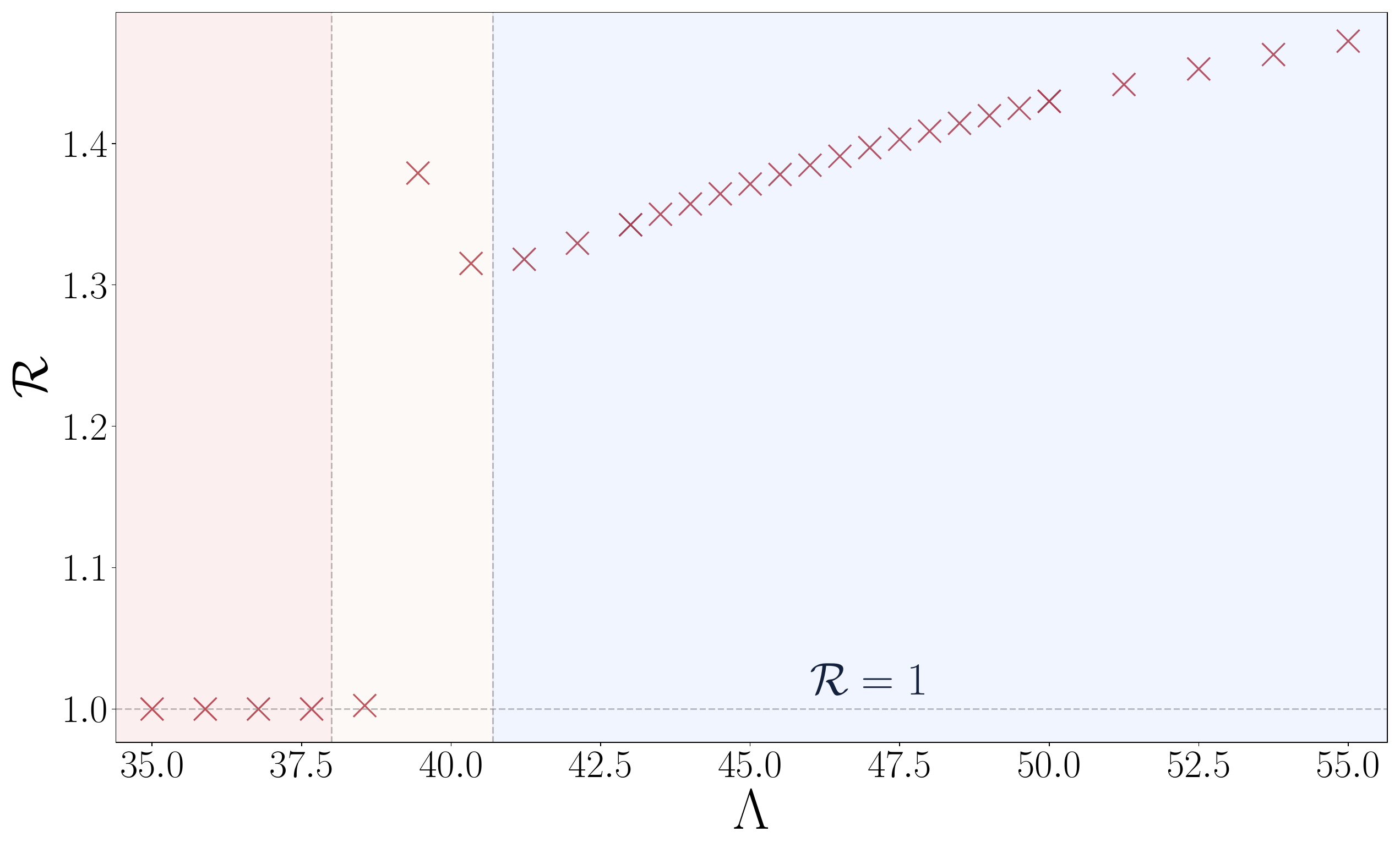}
    \caption{Compressibility ratio $\mathcal{R}$ for a system governed by a
    soft-core interaction of strength $\Lambda$. $\mathcal{R}$ is obtained using
    the relation \eqref{EQ: compressibility ratio definition}, where the values
    are calculated using homogenisation. The expected value of $\mathcal{R}=1$
    is obtained in the superfluid phase, but is starkly different in the
    supersolid regime.}
    \label{FIG:thermodynamic ratio}
\end{figure}
While we have shown that the compressibility ratio obtained via homogenisation
is in good agreement with the numerical determination, we do not yet fully
understand the nature of this anomalous result. In our search across the
literature we have not found another example of this phenomenon, and we are
currently investigating the origin of this effect. A possible explanation for
the result is considering a pressure which can be written like $P(n, G(n, V))$,
where $G$ is the ordering wavevector. One can then subsequently show that the
difference between bulk moduli can be written as
\begin{equation}
    B_e = B_t - V \frac{\partial P}{\partial G}\biggr|_n \frac{\partial G}{\partial V}\biggr|_n
\end{equation}
where it is clear that the last term is the anomalous term.

This term vanishes in ordinary materials in various ways. For gases and liquids,
there is no ordering vector, so this term never existed to begin with. A regular
solid has well-defined atomic positions. So, if the density is specified, the
ordering wavevector is also specified. Upon changing the volume, there is no
additional change in the ordering wavevector beyond that caused by the change in
density. Supersolids are unique in that the ordering vector $G$ experiences a
commensuration effect. For instance, in a one-dimensional system, it needs to be
an integer multiple of $2 \pi / L$ with $L$ being the length of the system. If
we stretch the system by $\delta L$ much less than the lattice spacing, then the
fractional change in the ordering wavevector is $\delta G/G = -\delta L/L$ so
that $L \partial G/\partial L = - G$. We believe this is a feature unique to
supersolids, due to their ability to exist at a wide range of densities, each of
which has their own ordering vector, whereas regular solids only have one
density at which they can exist (at some given pressure) and therefore only have
one ordering vector. This is (to the best of our knowledge) a new phenomena
which could be of great interest to the characterisation of supersolids.

Some currently open questions are as follows. Is this effect unique to the
supersolid phase, or is it a more general phenomena? What kind of thermal
transport is occurring in the phonon modes of a supersolid? Does the notion of
the Landau criterion for superfluidity still hold in the supersolid phase, and
if so, how does it change now that we have a band structure? These questions may
potentially have avenues to solution via the techniques developed in this paper.

The successful development of homogenisation techniques for the study of
supersolids is a significant step forward in the study of these systems. We have
shown that the homogenisation technique is able to accurately capture the
behaviour of the supersolid phase through comparison with well established
Bogoliubov techniques. We were able to obtain previously unknown elastic
properties of these materials as well as an effective Lagrangian which can be
used to study the dynamics of the system. In contrast with typical phases of
matter which have $\mathcal{R}=1$, we found that the compressibility ratio of
the supersolid phase is $\mathcal{R} > 1$. This is a new phenomena which we
predict will arise in supersolids, and of which the origin is currently unknown. 

\bibliography{supersolid.bib}

\clearpage 

\onecolumngrid
\appendix

\section{Expansion calculation}\label{app:lagrangian expansion}

\indent We start by expanding the non-interacting part of $\mathcal{L}$ using
the tools developed. We can express without expansion 
\begin{equation}
    \left( \nabla' \rho (\vec{r}') \right)^2 = \left[ \left( \partial_i + \left( u_{ik} + u_{il} u_{lk} \right) \partial_k \right) \left( \rho_0 (\vec{r}) + \tilde{\rho}(\vec{r}) \right) \right]^2
\end{equation}
where we subsequently drop the coordinates as it is clear everything is in the
unprimed basis $\mathbf{r}$. We can re-express the pre-factor
\begin{equation}
    \frac{1}{\rho (\vec{r}')} = \frac{1}{\rho_0 + \tilde{\rho}} = 
    \frac{1}{\rho_0} - \frac{\tilde{\rho}}{\rho_0^2} + \frac{\tilde{\rho}^2}{\rho_0^3} + \bigO(\tilde\rho^3) 
\end{equation} 
to make it more algebraically pliable. This allows us to write
\begin{equation}
    \begin{aligned} 
        \int_{\Omega'} \frac{\hbar^2}{2m} &\frac{(\nabla' \rho(\vec{r}'))^2}{4 \rho (\vec{r}')} \dif \vec{r}' 
        = \int_{\Omega} \Big\{ \frac{\hbar^2}{8m} \left( \frac{1}{\rho_0} - \frac{\tilde{\rho}}{\rho_0^2} + \frac{\tilde{\rho}^2}{\rho_0^3} \right)\left(1 + \epsilon_{ll} + M_{iklm} u_{ik} u_{lm} \right) \times\\
        &\left[(\nabla \rho_0)^2 + 2 (\grad \rho_0 \cdot \grad \tilde{\rho}) - 2 \epsilon_{ik} \partial_i \rho_0 \partial_k \rho_0 
        + 2 \chi_{ik} \partial_i \rho_0 \partial_k \rho_0 
        - 4 \epsilon_{ik} \partial_i \rho_0 \partial_k \tilde{\rho} + (\nabla \tilde{\rho})^2 
        + 2 \Delta_{ik} \partial_i \rho_0 \partial_k \rho_0  \right] \Big\} \dif \vec{r} 
    \end{aligned} 
\end{equation} 
which we can manipulate further. By isolating first and second order (in terms
of the strain tensor) parts of the non-interacting Lagrangian, and using the
following relation (obtained via integration by parts) 
\begin{equation}
    \int_\Omega \frac{\grad \tilde{\rho} \cdot \grad \rho_0}{\rho_0} \dif \vec{r} 
    = \int_\Omega \tilde{\rho} \frac{\left(\nabla \rho_0 \right)^2}{\rho_0^2}  - \tilde{\rho} \frac{\nabla^2 \rho_0 }{\rho_0} \dif \vec{r}
\end{equation} 
we are able to find the first-order component 
\begin{equation}
    \frac{\hbar^2}{4m} \left[ \left( \frac{(\nabla \rho_0)^2}{2 \rho_0^2} - \frac{\nabla^2 \rho_0 }{\rho_0} \right) \tilde{\rho} 
    + \epsilon_{ik} \left( \frac{(\nabla \rho_0)^2}{2 \rho_0} \delta_{ik} - \frac{\partial_i \rho_0 \partial_k \rho_0}{\rho_0} \right) \right] \\
\end{equation}
and the second-order component 
\begin{equation}
    \begin{aligned}
        & \frac{\hbar^2}{4m} \left[ \left( \Delta_{ik} + \chi_{ik} \right) \frac{\partial_i \rho_0 \partial_k \rho_0}{\rho_0} \right] + \\
        &\frac{\hbar^2}{4m} \frac{1}{2} \left[ -4 \epsilon_{ik} \frac{\partial_i \rho_0 \partial_k \tilde{\rho}}{\rho_0} \right.  + \frac{(\nabla \tilde{\rho})^2}{\rho_0} + 2 \frac{\grad \tilde{\rho} \cdot \grad \rho_0 }{\rho_0} \epsilon_{ll} 
         - 2 \epsilon_{ik} \frac{\partial_i \rho_0 \partial_k \rho_0}{\rho_0} \epsilon_{ll} \\
        &\qquad\quad + \frac{(\nabla \rho_0)^2}{\rho_0} \, M_{iklm} u_{ik} u_{lm} - 2 \frac{\grad \tilde{\rho} \grad \rho_0}{\rho_0^2} \tilde{\rho} 
        \left. + 2 \epsilon_{ik} \frac{\partial_i \rho_0 \partial_k \rho_0}{\rho_0^2}\tilde{\rho} 
        - \frac{(\nabla \rho_0)^2}{\rho_0^2}\tilde{\rho} \epsilon_{ll} + \frac{(\nabla \rho_0)^2}{\rho_0^3}\tilde{\rho}^2 \right] 
    \end{aligned}
\end{equation}
respectively. We now turn to considering the interaction section of the
Lagrangian. 

By first noting that the interaction potential is centrosymmetric (and indeed we
expect most effective potentials to be isotropic), we can write the interaction
as a function of radial distance. We can express the change in the metric length
of distance by simply applying the deformation definition to $\abs{\Delta
\vec{r}'}^2$ and obtain the result 
\begin{equation}
    \left( \Delta r' \right)^2 = \left( \delta_{ik} + 2 \epsilon_{ik} + 2 \Delta_{ik} \right) \left( \Delta r \right)_i \left( \Delta r \right)_k  
\end{equation}
which is exactly equivalent to that of Landau \& Lifshitz
\cite{landau1986theory}. The tensor in the above equation is sometimes referred
to as the finite strain tensor. 

Recalling the derivation of \eqref{EQ:f and W definitions} we define the
functions 
\begin{equation}
    U(\Delta \vec{r}') = U(\Delta \vec{r}) + \left( \epsilon_{ik} + \Delta_{ik} \right) f_{ik}(\Delta \vec{r}) + \epsilon_{ik}\epsilon_{lm} W_{iklm}(\Delta \vec{r})
\end{equation}
with 
\begin{equation}
    f_{ik}(\vec{r}) \equiv \frac{r_i r_k}{|\vec{r}|} \frac{\partial U (|\vec{r}|)}{\partial |\vec{r}|} 
    \quad ; \quad 
    W_{iklm}(\vec{r}) \equiv\frac{ r_i r_k r_l r_m}{2 |\vec{r}|^2}
    \left( \frac{\partial^2 U}{\partial |\vec{r^2}|} - \frac{1}{|\vec{r}|} \frac{\partial U}{\partial  |\vec{r}|} \right) \; .
\end{equation}

This can be used to express the interaction component of $\mathcal{L}$ by
combining it with both the density mapping and the Jacobian to obtain
\begin{equation}
    \begin{aligned}
        U = \frac{1}{2} \int_\Omega \left(\rho_0 + \tilde{\rho} \right)_{r_1} \left[ U(\vec{r}_{12}) 
        + \left( \epsilon_{ik} + \Delta_{ik} \right) f_{ik}( \vec{r}_{12}) \right.
        &\left.+ \epsilon_{ik}\epsilon_{lm} W_{iklm}(\vec{r}_{12}) \right] \left( \rho_0 + \tilde{\rho} \right)_{r_2}\\
        &\left(1 + 2\epsilon_{ll} + \epsilon_{ll}\epsilon_{kk} + 2 M_{iklm} u_{ik}u_{lm} \right)\; \dif \vec{r}_1 \dif\vec{r}_2 \quad 
    \end{aligned}
\end{equation}
where $\vec{r}_{12} = \vec{r}_1-\vec{r}_2$. We note that, since the convolution
is integrated, there is a term arising from $\left( \mathcal{J}_{\vec{r}'
\rightarrow \vec{r}} \right)^2$. 

Once again, this can be expanded and combined with the non-interacting component
of $\mathcal{L}$, and we can fully express the first and second order components
of the expansion. We make use of the Gross-Pitaevskii ground-state equation \eqref{EQ:rho base EoM}, as well as the normalisation
condition \eqref{EQ:strained normalisation condition} to simplify our
expressions, and we can show that 
\begin{equation}\label{linear terms simplified}
    \mathcal{L}_1 = 
    \int_\Omega \left[\rho_0 \epsilon_{ll} \left( U*\rho_0 - \mu_0 \right)
     + \frac{1}{2} \epsilon_{ik} (f_{ik} * \rho_0)\rho_0  
     + \frac{\hbar^2}{4m} \epsilon_{ik} \left(
        \frac{\left( \nabla \rho_0 \right)^2}{2 \rho_0} \delta_{ik} 
        - \frac{\partial_i \rho_0 \partial_k \rho_0}{\rho_0} 
    \right)\right] \dif \vec{r }
\end{equation}
We can now make a direct connection between the linear component of the
expansion and the pressure tensor $P_{ik}$. Recalling that the first order
component of the expansion is the stress tensor, and that the stress tensor is
the negative of the pressure, we can write 
\begin{equation}\label{pressure equation}
\begin{split}
    P_{ik} &= \frac{1}{|\Omega|} \int_\Omega \left[\left( \mu_0 - U * \rho_0 \right) \delta_{ik} 
    - \frac{1}{2} f_{ik} * \rho_0 
    - \frac{\hbar^2}{4m} \left(
        \frac{\left( \nabla \rho_0 \right)^2}{2 \rho_0^2} \delta_{ik} 
        - \frac{\partial_i \rho_0 \partial_k \rho_0}{\rho_0^2} \right)
    \right] \rho_0 \dif \vec{r} \\
    &= \frac{1}{|\Omega|} \int_\Omega \left[ \frac{\hbar^2}{4m}\left(
     \frac{\partial_i \rho_0 \partial_k \rho_0}{\rho_0} - \nabla^2\rho_0 \,\delta_{ik}\right)
     - \frac{1}{2}\rho_0 (f_{ik} * \rho_0)  \right]\dif \vec{r} \\
    \end{split}
\end{equation}
The second line consists of the ground-state density profile only.
This is our first key result which is obtaining the pressure of any ground-state
system given a specified interaction. We can trivially calculate the expected
pressure for a homogeneous superfluid by noting that $\rho_0 \equiv n$ and that $\mu_0 = U_{\vec{k}=0}n$ where $ U_{\vec{k}=0} = \int_\Omega U(\vec{r})d\vec{r}$.
The pressure simply reduces to $-\frac{1}{2} n^2 \int_{\Omega} f_{ik} (\vec{r})
\dif \vec{r}$. Integrating by parts gives the integral over $f_{ik}$ as $-U_{\vec{k}=0} \delta_{ik}$ . 
So, we find that $P_{ik} = \frac{1}{2}\mu_0 n \delta_{ik}$, which is the expected
result. We can now move on to the more difficult task of solving for the second
order components. 

It is important to note that since we have allowed $\tilde{\rho}$ to have both
first and second order components in $u_{ik}$, then even single powers of
$\tilde{\rho}$ must be taken into consideration for the $\bigO(u^2)$ components.
The above analysis only collected the components of $\tilde{\rho}$ which were of
$\bigO(u)$, but does not remove the singular powers of $\tilde{\rho}$ from the
Lagrangian. It is necessary to calculate the functional form of $\tilde{\rho}$
as there will be terms like $\int_\Omega \tilde{\rho}^2 \dif \vec{r}$ that will
contribute to the perturbative Lagrangian, and as such we therefore need to
solve the equations of motion for $\tilde{\rho}$. It is not good enough to
simply know the normalisation condition as we did for the $\bigO(u)$ component.

The first step in solving for $\tilde{\rho}$ is to collect all the terms in
which it appears, regardless of first or second order splitting. We take all the
terms that contain $\tilde{\rho}$ and call them $\mathcal{L}_{\tilde{\rho}}$,
finding
\begin{equation}\label{Lagrangian of rho_tilde}
    \begin{aligned}
        - \mathcal{L}_{\tilde{\rho}} = & \frac{1}{2} \frac{\hbar^2}{4m} \left[ -4 \epsilon_{ik} \frac{\partial_i \rho_0 \partial_k \tilde{\rho}}{\rho_0}
        + \frac{(\nabla \tilde{\rho})^2}{\rho_0} 
        + 2 \frac{\grad \tilde{\rho} \cdot \grad \rho_0}{\rho_0} \epsilon_{ll} - 2 \frac{\grad \tilde{\rho} \cdot \grad \rho_0}{\rho_0^2} \tilde{\rho} 
        + 2 \epsilon_{ik} \frac{\partial_i \rho_0 \partial_k \rho_0}{\rho_0^2} \tilde{\rho} 
        - \frac{(\nabla \rho_0)^2}{\rho_0^2} \tilde{\rho} \, \epsilon_{ll} + \frac{(\nabla \rho_0)^2}{\rho_0^3} \tilde{\rho}^2 \right] \\ 
        & + \mu_0 \tilde{\rho} + \frac{1}{2} \left( \left(U * \tilde{\rho} \right) \tilde{\rho} + 4 \epsilon_{ll} \left( U * \rho_0 \right) \tilde{\rho} + 2 \epsilon_{ik} \left(f_{ik} * \rho_0 \right) \tilde{\rho} \right) 
    \end{aligned}
\end{equation}
which we can solve by the standard Euler-Lagrange technique. A key detail now is
that the new normalisation condition changes the constraints we apply. Specifically, the equation we need to solve is
\begin{equation}
\frac{\partial \mathcal{L}_{\tilde{\rho}}}{\partial \tilde{\rho}} 
- \partial_i \frac{\partial \mathcal{L}_{\tilde{\rho}}}{\partial \left( \partial_i \tilde{\rho} \right)}
+ \left( \mu_0 + \mu_1 \epsilon_{ll} + \mu_2^{iklm} u_{ik} u_{lm} \right) 
   \left( 1 + \epsilon_{ll} + M_{iklm} u_{ik} u_{lm} \right) = 0    
\end{equation}
which can be shown via the strained normalisation condition \eqref{EQ:strained
normalisation condition}. 

A straightforward calculation follows of differentiating the $\tilde{\rho}$
dependent Lagrangian and collating all terms. We make use of the zeroth-order
equation of motion to simplify our expression and we find 
\begin{equation}\label{app:full EoM in rho_tilde}
    \begin{split}
        &\frac{\hbar^2}{4m} \grad \cdot \left( \frac{\grad \tilde{\rho}}{\rho_0} \right) - U * \tilde{\rho} \\[5pt]
        + &\frac{\hbar^2}{4m} \left( \frac{(\nabla \rho_0)^2}{\rho_0^3} - \frac{\nabla^2 \rho_0}{\rho_0^2} \right) \tilde{\rho}
        \end{split}
        \text{\Large$ = $} 
        \begin{split}
        &- \left( \mu_2^{iklm} + \mu_1^{ik} \delta_{lm} + \mu_0 M_{iklm} \right) u_{ik} u_{lm} 
        + \epsilon_{ik} \left[ -  \mu_1^{ik} + \frac{\hbar^2}{4m} \left( 2 \frac{\partial_{ik} \rho_0}{\rho_0}
        - \frac{\partial_i \rho_0 \partial_k \rho_0}{\rho_0^2} \right) \right. \\[5pt]
        & \left. + \int_\Omega \left( f_{ik}(\vec{r} - \vec{r}') + \delta_{ik} U (\vec{r} - \vec{r}') \right) \rho_0 (\vec{r}') \dif \vec{r}'
        \right]
    \end{split}
\end{equation}
which contains both $\bigO(u^2)$ and $\bigO(u)$ terms. We can now begin to
separate $\tilde{\rho}$ into first and second order components via \eqref{EQ:rho
tilde expansion strain}, and consider solving the equation component by
component. That is to say that we solve the lowest order component first, and
then use that solution to solve the next order component, and so on.

The first order equation to solve is given by 
\begin{equation}\label{rho_1 equation of motion}
    \begin{split}
        & \frac{\hbar^2}{4m} \grad \cdot \left( \frac{\grad \left( \rho_1^{ik} \epsilon_{ik} \right)}{\rho_0} \right) - U * \left( \rho_1^{ik} \epsilon_{ik} \right) \\
        & + \frac{\hbar^2}{4m} \left( \frac{(\nabla \rho_0)^2}{\rho_0^3} - \frac{\nabla^2 \rho_0}{\rho_0^2} \right) \left( \rho_1^{ik} \epsilon_{ik} \right)  
    \end{split}
    \text{\Large$ = $}
    \begin{split}
        \epsilon_{ik} & \left[ -  \mu_1^{ik} + \frac{\hbar^2}{4m} \left( 2 \frac{\partial_{ik} \rho_0}{\rho_0}
        - \frac{\partial_i \rho_0 \partial_k \rho_0}{\rho_0^2} \right) \right. \\[5pt]
        & \left. + \int_\Omega \left( f_{ik}(\vec{r} - \vec{r}') + \delta_{ik} U (\vec{r} - \vec{r}') \right) \rho_0 (\vec{r}') \dif \vec{r}'
        \right]
    \end{split}
\end{equation}
which we can solve numerically. We factor out $\epsilon_{ik}$ and solve the set
of equations where we have one for each index (in 2D we have 4 equations to
solve, in 3D there are 9). As this is an integrodifferential equation we can
solve the equation by writing the LHS as a matrix operator on an unraveled
vector and applying an iterative matrix solver algorithm which converges on some
pre-specified tolerance. We can introduce notation and write the first and
second order equations of motion as 
\begin{equation}\label{equation of motion shortened notation}
        L \left( \rho_1^{ik} \right) = F_1 \left( \rho_0, \mu_1^{ik} \right) \,, \qquad 
        L \left( \rho_2^{iklm} \right) = F_2 \left( \mu_0, \mu_1^{ik}, \mu_2^{iklm} \right) 
\end{equation}
for brevity. It is important to note that we do not actually \textbf{need} to
solve the equation for $\rho_2^{iklm}$ as it only appears in the Lagrangian when
multiplied by a constant, thus we can directly use the normalisation condition
\eqref{EQ:strained normalisation condition} and short circuit the task of
solving it. We also do not need to find $\mu_2^{iklm}$ for a slightly different
reason, which is that it appears in the Lagrangian to too high order and so is
always discarded anyway. Thus our only task now is to determine the first order
correction to the chemical potential $\mu_1^{ik}$ and the first order solution
to $\tilde{\rho}$ which is $\rho_1^{ik}$. 

We first state $\rho_1^{ik} \equiv \rho_1^{ik} ( \vec{r}, \mu_1^{ik})$, i.e. it
is now a function of the chemical potential. We specifically choose the
functional form $\rho_1^{ik} = \hat{\rho}_1^{ik} + \mu_1^{ik} \rho_1'$, in a
kind of 'homogeneous + inhomogeneous' fashion. Namely, we demand that 
\begin{equation}
        L \left( \hat{\rho}_1^{ik} \right) = \frac{\hbar^2}{4m} \left( 2 \frac{\partial_{ik} \rho_0}{\rho_0}
        - \frac{\partial_i \rho_0 \partial_k \rho_0}{\rho_0^2} \right)
        + \int_\Omega \left[ f_{ik}(\vec{r} - \vec{r}') + \delta_{ik} U (\vec{r} - \vec{r}') \right] \rho_0 (\vec{r}') \dif \vec{r}'    
\end{equation}
i.e. $\hat{\rho}_1^{ik}$ satisfies the equation of motion without any chemical
potential considerations. Then we solve the equation 
\begin{equation}
    L \left( \mu_1^{ik} \rho_1' \right) = - \mu_1^{ik} \, \rightarrow \, L \left( \rho_1' \right) = -1
\end{equation}
simultaneously, which we can do numerically as it is the same operator but just
a different RHS. We can state this in the notation of \eqref{equation of motion
shortened notation}
\begin{equation}
    L \left( \hat{\rho}_1^{ik} \right) = F_1 \left( \rho_0, \mu_1^{ik} = 0 \right) \quad ,\quad 
    L \left( \rho_1' \right) = F_1 \left( \rho_0 = 0, \mu_1^{ik} = 1 \right) 
\end{equation}
which we can readily substitute into the 'homogeneous' and 'inhomogeneous'
solutions to see that 
\begin{equation}
    L \left( \hat{\rho}_1^{ik} + \mu_1^{ik} \rho_1' \right) = F_1 \left( \rho_0, \mu_1^{ik} \right) 
\end{equation}
which is the `general' solution we were looking for. The final task is to now
actually calculate $\mu_1^{ik}$ which we can do by using the normalisation
condition\eqref{EQ:strained normalisation condition}. Since we know what the first-order condition is, we can rearrange to find that 
\begin{equation}\label{mu_1^ik solution}
    \mu_1^{ik} = - \int_{\Omega} (\rho_0 \delta_{ik} + \hat{\rho}_1^{ik}) \dif \vec{r} \Bigg/ \int_\Omega \rho_1' \dif \vec{r}
\end{equation}
which is the final ingredient. 

We have now fully solved for $\tilde{\rho}$ (and discarded the parts that we
don't need), so we can re-express $\mathcal{L}_{\tilde{\rho}}$ using integration
by parts. The calculation is relatively straightforward and makes use of the
base EoM \eqref{EQ:rho base EoM}, allowing one to find 
\begin{equation}\label{final lagrangian of rho_tilde}
    - \mathcal{L}_{\tilde{\rho}} = \frac{1}{2} \left( 2 \mu_0 \rho_2^{iklm} u_{ik} u_{lm} 
    + \rho_1^{ik} \left[ 
    \frac{\hbar^2}{4m} \left( 2 \frac{\partial_{ik} \rho_0}{\rho_0} - \frac{\partial_i \rho_0 \partial_k \rho_0}{\rho_0^2} \right)
    + \left( f_{ik} + \delta_{ik} U \right) * \rho_0 + 2 \mu_0 \delta_{ik} + \mu_1^{ik} \right] \, \epsilon_{ik} \epsilon_{lm}
    \right) + \bigO(\epsilon^3) 
\end{equation}
which contains all the second-order contributions from $\tilde{\rho}$. 

We can combine this with all the second-order contributions that \textbf{do not
contain} $\tilde{\rho}$, and obtain a long expression 
\begin{equation}\label{app:quadratic lagrangian density}
    \begin{split}
        - \mathcal{L}_2 = \frac{1}{2} & \left[
        2 \left( U * \rho_0 \right) \rho_0 M_{iklm} u_{ik} u_{lm} 
        + \left( \left( U * \rho_0 \right) \rho_0 \delta_{ik} \delta_{lm} + 2 \left( f_{ik} * \rho_0 \right) \rho_0 \delta_{lm} 
        + \left( W_{iklm} * \rho_0 \right) \rho_0 \right) \, \epsilon_{ik} \epsilon_{lm} \right. \\
        & \left. 
        + \left( f_{ik} * \rho_0 \right) \rho_0 \Delta_{ik} + 2 \frac{\hbar^2}{4m} \frac{\partial_i \rho_0 \partial_k \rho_0}{\rho_0} \left(\Delta_{ik} + \chi_{ik} \right) 
        + \frac{\hbar^2}{4m} \left( -2 \frac{\partial_i \rho_0 \partial_k \rho_0}{\rho_0} \right) \delta_{lm} \epsilon_{ik} \epsilon_{lm} \right. \\ 
        & \left. 
        + \frac{\hbar^2}{4m} \frac{\left( \nabla \rho_0 \right)^2}{\rho_0} M_{iklm} u_{ik} u_{lm} + 2 \mu_0 \rho_0 \left( \delta_{ik} \delta_{lm} - M_{iklm} \right) u_{ik} u_{lm} \right. \\
        & \left. 
        + \rho_1^{lm} \left[ 
        \frac{\hbar^2}{4m} \left( 2 \frac{\partial_{ik} \rho_0}{\rho_0} - \frac{\partial_i \rho_0 \partial_k \rho_0}{\rho_0^2} \right)
        + \left( f_{ik} + \delta_{ik} U \right) * \rho_0 + 2 \mu_0 \delta_{ik} + \mu_1^{ik} \right] \, \epsilon_{ik} \epsilon_{lm} 
        \right] \,.
    \end{split}
\end{equation}
which will be used in the calculation of the elastic constants. 

To summarise, we have expanded our Lagrangian up to second order in the strain
tensor and collected all the terms. We've solved for $\tilde{\rho}$ such that
the new state is a ground state with the strained normalisation condition, and
subsequently re-expressed the second-order Lagrangian with that solution in
hand. 

\section{Connecting Cauchy elastic tensor to expansion}\label{app:Bavaud}
\twocolumngrid

This section closely follows the calculation done by Bavaud et al
\cite{bavaud1986statistical}. To begin with, we define an analytic expansion in
the free energy of our system around a deformation $x_{\alpha}' = \left(
\delta_{\alpha \beta} - u_{\alpha \beta} \right) x_{\beta}$ with a mapping
$\Omega \mapsto \Omega'$ 
\begin{equation}\label{free energy bavaud expansion}
    F( \Omega' ) = F ( \Omega ) - \abs{\Omega} \pi_{\alpha \beta} ( \Omega ) u_{\alpha \beta} 
    + \abs{\Omega} \frac{1}{2} A_{\alpha \beta \gamma \delta} ( \Omega ) u_{\alpha \beta} u_{\gamma \delta} 
\end{equation}
where 
\begin{equation}\label{pi_ab definition}
    \pi_{\alpha \beta} ( \Omega ) = \left. \frac{1}{\abs{\Omega}} \frac{\partial F (\Omega') }{\partial \left( - u_{\alpha \beta} \right)} \right|_{u=0} 
    = \left. - \frac{1}{\abs{\Omega}} \frac{\partial F ( \Omega' )}{\partial u_{\alpha \beta}} \right|_{u=0} 
\end{equation}
and similarly
\begin{equation}\label{A_iklm definition} 
    A_{\alpha \beta \gamma \delta} (\Omega) = \left. \frac{1}{\abs{\Omega}} \frac{\partial^2 F ( \Omega' )}{\partial u_{\alpha \beta} u_{\gamma \delta}} \right|_{u=0} \quad. 
\end{equation}
To clarify, we map between spaces $\Omega \mapsto \Omega'$ through the
transformation $\vec{x} \mapsto \vec{x'}$. 

It is important to note that the tensor $A_{\alpha \beta \gamma \delta}$ is
\textbf{not} the elastic tensor. It is the expansion of the free energy due to
the deformation, but the elastic tensor strictly connects the stress tensor to
the strain. We are searching for Hooke's Law which maps strain in a deformed
point to strain in an undeformed point but now with the notation
\begin{equation}\label{Hooke's law Bavaud} 
    \pi_{ik} ( \Omega') = \pi_{ik} (\Omega) - B_{iklm} ( \Omega ) u_{lm} \quad .
\end{equation}
An important additional feature of the elastic tensor $B_{iklm}$ is that, for a
sufficiently symmetric system, it has certain symmetry properties as described
by Landau \& Lifshitz \cite{landau1986theory}, whereas the expansion tensor
$A_{iklm}$ \textbf{does not} necessarily have these symmetries. 

We first define an additional two deformations, $u'$ and $u''$ such that
\begin{equation}
    x_{\alpha}'' =  \left( \delta_{\alpha \beta} - u_{\alpha \beta}' \right) x_{\beta}' \quad ; \quad 
    x_{\alpha}'' = \left( \delta_{\alpha \beta} - u_{\alpha \beta}'' \right) x_{\beta} 
\end{equation}
To clarify, we have an initial space $\Omega$ which we can deform with the
transformation $\vec{u}$ into the space $\Omega'$. We can then deform from
$\Omega' \mapsto \Omega''$ with the deformation $\vec{u}'$, and we can
\textit{also} deform to the $\Omega''$ space from the \textit{original space}
like $\Omega \mapsto \Omega''$ with the deformation $\vec{u}''$. We can combine
the deformations to find the relation 
\begin{equation}\label{u'' relation} 
    u_{\alpha \beta}'' = u_{\alpha \beta} + u_{\alpha \beta}' - u_{\alpha \sigma}' u_{\sigma \beta} \quad. 
\end{equation}
We then use the above relation to expand the stress tensor that connects the
$\Omega'$ and $\Omega''$ space. Examining the expansion of the free energy from
the $\Omega'$ space to the $\Omega''$ space, we find 
\begin{equation}
    F( \Omega'') = F ( \Omega' ) - \abs{\Omega'} \pi_{\alpha \beta} ( \Omega' ) u_{\alpha \beta}' 
    + \frac{\abs{\Omega'} }{2} A_{\alpha \beta \gamma \delta} ( \Omega' ) u_{\alpha \beta}' u_{\gamma \delta}' \,.  
\end{equation}
Expanding $\pi_{\alpha \beta} ( \Omega' )$ and using \eqref{pi_ab definition} and \eqref{u'' relation}, we find
\begin{equation}\label{pi_ab partial derivation}
    \begin{split}
        \pi_{\alpha \beta} \left( \Omega' \right) 
        =& \left. - \frac{1}{\abs{\Omega'}} \frac{\partial F (\Omega'')}{\partial u_{\alpha \beta}'} \right|_{u'=0} \\
        =& \left. - \frac{1}{\abs{\Omega'}} \frac{\partial F (\Omega'')}{\partial u_{\gamma \epsilon}''} 
        \frac{\partial u_{\gamma \epsilon}''}{\partial u_{\alpha \beta}'} \right|_{u'=0} \\
        =& \left. - \frac{1}{\abs{\Omega'} } \frac{\partial F( \Omega'')}{\partial u_{\gamma \epsilon}''} 
        \left( \delta_{\gamma \alpha} \delta_{\beta \epsilon} - \delta_{\gamma \alpha} \delta_{\beta \sigma} u_{\sigma \epsilon} \right) \right|_{u' = 0} \\
        =& \left. - \frac{1}{\abs{\Omega'}} \left( \delta_{\beta \epsilon} - u_{\beta \epsilon} \right) \frac{\partial F( \Omega'' )}{\partial u_{\alpha \epsilon}''} \right|_{u'=0} \,. 
    \end{split}
\end{equation}
We can now turn our attention to the derivative in the above expression. Namely,
since the derivative actually does not depend on $u'$ explicitly (recall that we
can map from $\Omega \mapsto \Omega''$ through $u''$ without needing $u'$ at
all), we can remove the evalutation at $u'=0$. As such, the derivative is now
not a number, but a function instead! This function describes the energy
response of the mapping $\Omega \mapsto \Omega''$ through $u''$ for any
arbitrary $\Omega''$ and $u''$. That means it \textbf{must be equivalent} to the
function which maps from $\Omega \mapsto \Omega'$ through $u$. Thus we can write
\begin{equation}
    \left. \frac{\partial F( \Omega'')}{\partial u_{\alpha \epsilon}''} \right|_{u'=0}
    = \frac{\partial F( \Omega')}{\partial u_{\alpha \epsilon}}
\end{equation}
and the final result from \eqref{pi_ab partial derivation} becomes 
\begin{equation}\label{pi_ab intermediate result} 
    \pi_{\alpha \beta} \left( \Omega' \right) = 
    - \frac{1}{\abs{\Omega'}} \left( \delta_{\beta \epsilon} - u_{\beta \epsilon} \right) \frac{\partial F ( \Omega' )}{\partial u_{\alpha \epsilon}} \quad. 
\end{equation}
Since the derivative is no longer evaluated at $u=0$, we now substitute in
\eqref{free energy bavaud expansion} to find 
\begin{equation}\label{free energy derivative bavaud}
    \frac{\partial F(\Omega')}{\partial u_{\alpha \epsilon}} = 
    - \abs{\Omega} \pi_{\alpha \epsilon} ( \Omega) + \abs{\Omega} A_{\alpha \epsilon \gamma \delta} ( \Omega ) u_{\gamma \delta} \quad . 
\end{equation}
We now also expand the $1/ \abs{\Omega'}$ term to find 
\begin{equation}\label{bavaud volume expansion}
    \frac{1}{\abs{\Omega'}} = \frac{1}{\abs{\Omega}} \frac{1}{\det(\mathds{1} - \vec{U})}
    = \frac{1}{\abs{\Omega}} \left( 1 + \delta_{x y} u_{x y} + \bigO(2) \right) 
\end{equation}
where the indices $x$ and $y$ are tensor indices of the same nature as $\alpha$,
$\beta$ etc. It is important to note that we are discarding terms of
$\bigO(u^2)$ or higher as we are trying to derive Hooke's law, i.e. $F=kx$, and
we do not want anharmonic terms. 

We can now combine the results of \eqref{free energy derivative bavaud} and
\eqref{bavaud volume expansion} into \eqref{pi_ab intermediate result}
(discarding $\bigO(u_{ik}^2)$ terms) to find 
\begin{equation}
    \pi_{\alpha \beta} ( \Omega' ) = \pi_{\alpha \beta} - \left(A_{\alpha \beta \gamma \delta}   
    + \delta_{\gamma \beta} \pi_{\alpha \delta} - \delta_{\gamma \delta} \pi_{\alpha \beta} \right) u_{\gamma \delta}
\end{equation}
where everything on the RHS is a function of $\Omega$. This is the expression of
Hooke's Law as described in \eqref{Hooke's law Bavaud}! We can therefore find
that 
\begin{equation}
    B_{iklm} = A_{iklm} + \pi_{im} \delta_{kl} - \pi_{ik} \delta_{lm} 
\end{equation}
with now the final task being to relate $\pi_{ik}$ to the pressure tensor
$P_{ik}$. Recalling that the stress tensor is just the negative of the pressure,
we can write the final expression as 
\begin{equation}\label{app:stress-strain tensor} 
    B_{iklm} = A_{iklm} + P_{ik}\delta_{lm} - P_{im} \delta_{kl} 
\end{equation}
which is the Cauchy elastic tensor. For the case of $C_3$ symmetry, 
we have the relation \cite{landau1986theory}: 
\begin{equation}\label{app:Brelation} 
    B_{xxxx} = B_{xxyy} + 2B_{xyxy}\,.
\end{equation}

\section{Tensor differentiation}
\label{app:derivative tensors}

We have shown in previous sections how to expand a Lagrangian under a strain tensor. Once we have that expansion, how do we recover the homogenised tensors as first and second differentives: $\pi_{ik} = \partial F/\partial u_{ik}$ and $A_{iklm} =
\partial^2 F/\partial u_{ik} \partial u_{lm}$?

We will now write out explicitly how expressions involving the tensors listed in
\eqref{brevity tensor definitions} differentiate.
Consider first $F = F_{ik} u_{ik}$ with $F_{ik} = F_{ki}$, its first derivative is:
$\partial F/\partial u{ik} = F_{ik}$.
Similarly, the first derivative of $F = F_{ik} \epsilon_{ik}$ is 
$\partial F/\partial u_{ik} = -F_{ik}$.

\begin{widetext}
Let us now consider second derivatives. Suppose there is a term in the Lagrangian of the form $F = F_{iklm} u_{ik} u_{lm}$. We have the result $\partial^2 F/\partial u_{ik} \partial u_{lm}  = 2F_{iklm}$. 
Consider now $F = F_{ik} \chi_{ik}$ where 
$\chi_{ik}=\left( u_{li} u_{kl} + u_{lk} u_{il} \right)/2$. Since $\chi_{ik}$ is symmetric under $i \leftrightarrow k$, we can impose that $F_{ik}$ is also symmetric under the same
exchange. 
\begin{equation}
    \begin{aligned}
      \frac{\partial^2 \chi_{ik}}{\partial u_{cd} \partial u_{ab}}   &= \frac{1}{2} \frac{\partial}{\partial u_{cd}} \left( \delta_{la} \delta_{ib} u_{kl} + \delta_{la} \delta_{kb} u_{il}
        + \delta_{ka} \delta_{lb} u_{li} + \delta_{ia} \delta_{lb} u_{lk} \right) 
        = \frac{1}{2} \left( \delta_{ib} \delta_{kc} \delta_{ad} + \delta_{kb} \delta_{ci} \delta_{ad} + \delta_{ka} \delta_{bc} \delta_{id} + \delta_{ia} \delta_{bc} \delta_{kd} \right)\,.
    \end{aligned}
\end{equation}
This allows us to obtain
\begin{equation}
    \frac{\partial^2 F}{\partial u_{ik} \partial u_{lm}} = \frac{1}{2} \left( F_{bc} \delta_{ad} + F_{cb} \delta_{ad} + F_{da} \delta_{bc} + F_{ad} \delta_{bc} \right) 
    = F_{kl} \delta_{im} + F_{im} \delta_{kl} \,.
\end{equation}
using the symmetry of $F_{ik}$ under exchange of indices. A similar calculation can be done for other tensor contractions involving the tensors in \eqref{brevity tensor definitions}. This is listed in Table \ref{tab:tensorderivs}.

\begin{table}[htb]
\caption{\label{tab:tensorderivs}}
\begin{tabular*}{0.5\textwidth}{@{\extracolsep{\fill}}|c||l|l|l|l|}
\hline
    $F$ &$F_{iklm} \epsilon_{ik} \epsilon_{lm}$ & $F_{ik} \Delta_{ik}$ & $F_{ik}\omega_{ik}$ & $F_{ik}\chi_{ik}$\\ \hline
    $\displaystyle{\frac{\partial^2 F}{\partial u_{ik}\partial u_{lm}}}$
    & $2F_{iklm}$ 
    & $F_{il} \delta_{km}$
    & $F_{km} \delta_{il}$ 
    & $F_{kl} \delta_{im} + F_{im} \delta_{kl}$\\ \hline
\end{tabular*}
\end{table}

It is important to briefly discuss the symmetry of each of these expressions.
For terms that look like $F_{iklm} \epsilon_{ik} \epsilon_{lm}$, these already
have all the symmetry properties we would want from our elastic tensors.
However, we can immediately see that the contribution from $F_{ik} \Delta_{ik}$,
$F_{ik} \omega_{ik}$ and $F_{ik} \chi_{ik}$ \textbf{do not} have the requisite
symmetries. We instead need to \textit{create} these symmetries from
combinations of these tensors. This cannot be achieved a priori by adding in
whatever we like. However, if we expect elastic theory to hold, then we would expect
any expansion in any strain tensor (including ours) to lead us eventually back to symmetric expressions. This is a good test to check whether the expansion calculation has been done correctly.
\end{widetext}

\section{Numerical Deformations}\label{app:numerical elastics}

In our numerics, we use a spatial discretisation consisting of a triangular lattice as a grid so that the numerical scheme is consistent with the symmetry of the supersolid phase under rotations. Therefore, each ``pixel" is a parallelogram and the simulation cell is also a parallelogram with basis vectors $\vec{a}$ and $\vec{b}$ parallel to $(\sqrt{3}/2,1/2)$ and $(0,1)$ in Cartesian coordinates. This appendix provides some algebraic details on how deformations $\vec{V}$ of the simulation cell can be used to extract the elastic tensors numerically using the energy change $F_2 \equiv P u_{ll} + \frac{1}{2} A_{iklm} u_{ik} u_{lm}$ up to second order in the strain.

Let us write all positions and displacements in the basis of the triangular grid: $\vec{r} = r_a \vec{a} + r_b \vec{b}$ and $\vec{u} = u_a \vec{a} + u_b\vec{b}$. The strain tensor in this basis
\[
\vec{V} = \begin{pmatrix}
\partial u_a/\partial r_a & 
\partial u_a/\partial r_b \\
\partial u_b/\partial r_a & 
\partial u_b/\partial r_b 
\end{pmatrix}
\]  
is related to the strain tensor in the Cartesian basis: $(\vec{U})_{ij} \equiv u_{ij}$ by:
\begin{equation}
\vec{U} = \vec{T}^{-1} \vec{V} \vec{T}\,,\qquad
\mbox{with}\qquad
    \vec{T} = 
    \begin{pmatrix}
        \frac{2}{\sqrt{3}} & 0 \\
        - \frac{1}{\sqrt{3}} & 1 
    \end{pmatrix}\,.
\end{equation}

In our numerics, we apply strains diagonal in the triangular-grid basis by deforming the simulation cell in the directions of $\vec{a}$ and $\vec{b}$. Using the above, we can calculate the energy changes we should expect to see. There are three deformations that give linearly independent results.
\begin{align}\label{list of free energy expansions due to parallelogram strain}
        \vec{V} = 
        \begin{pmatrix}
            \epsilon & 0 \\ 
            0 & 0 
        \end{pmatrix}\!:
        F_{2} &=  P \epsilon +\left( A_{xxxx} + \frac{A_{xyxy}}{3}  \right) \frac{\epsilon^2}{2} \notag\\ 
        \vec{V} = 
        \begin{pmatrix}
            0 & 0 \\ 
            0 & \epsilon 
        \end{pmatrix}\!:
        F_{2} &= P \epsilon + \left( A_{yyyy} + \frac{A_{xyxy}}{3}  \right) \frac{\epsilon^2}{2} \\ 
        \vec{V} = 
        \begin{pmatrix}
            \epsilon & 0 \\ 
            0 & \epsilon 
        \end{pmatrix}\!:
        F_{2} &= 2P \epsilon +\left( \frac{A_{xxxx}+ A_{yyyy}}{2} + A_{xxyy} \right) \epsilon^2\,. \notag
\end{align}

For our system with $C_3$ symmetry, there are only two independent elastic constants. Using \eqref{app:stress-strain tensor} and
\eqref{app:Brelation}, we see that $A_{xxxx} = A_{xxyy} + P + 2 A_{xyxy}$. Therefore, we have three equations to solve for the elastic constants and the pressure for a given $\epsilon \ll 1$.

We have calculated these quantities at several values of $\epsilon$ to check for the self-consistency of this analysis. This gives a measure of the numerical errors of this technique.

A good question to ask is whether or not we are allowed to \textit{a priori}
assume the system is symmetric, and whether we are biasing towards a symmetric
result. This is a valid question, but really what we are doing is seeking to
compare the results we obtain with the analytical theory derived. Importantly,
the theory assumes nothing about symmetry, so if the numerics agree with the
theory, then this assumption is justified. 

\end{document}